\documentclass[journal]{IEEEtran}

\usepackage[english]{babel}

\usepackage{ifpdf}

\usepackage{cite} 
\usepackage{url}
\usepackage{hyperref}

\ifCLASSINFOpdf
	\usepackage[pdftex]{graphicx}
	\graphicspath{{./figures/}}
\else
	\usepackage[dvips]{graphicx}
	\graphicspath{./figures/}
\fi
\usepackage{color}
\usepackage{pgf, tikz, pgfplots}
\usepackage{tikz-network}
\usetikzlibrary{shapes, arrows, automata, plotmarks}
\usetikzlibrary{calc,hobby,decorations}

\usepackage[safe]{tipa}
\usepackage{amsmath}
\usepackage{amsfonts, amssymb, amsthm}
\usepackage{mathrsfs}
\usepackage{dsfont}
\usepackage{orcidlink}

\usepackage{algorithm,algpseudocode}
\algnewcommand{\LeftComment}[1]{\Statex \(\triangleright\) #1}

\usepackage{enumerate}
\usepackage{multirow}
\usepackage{rotating}
\usepackage{breqn}
\usepackage{soul}
\usepackage{subcaption}
\usepackage{pgfplots}
\usepgfplotslibrary{groupplots,fillbetween}
\usetikzlibrary{calc}
\pgfplotsset{compat=1.18}

\usepackage{needspace}

\usepackage{amsmath}

\input{./latex_resources/mySymbol.sty}
\input{./latex_resources/asp_colors.sty}

\newtheorem{lemma}{\hspace{0pt}\bf Lemma}
\newtheorem{proposition}{\hspace{0pt}\bf Proposition}

\newtheorem{theorem}{\hspace{0pt}\bf Theorem}
\newtheorem{corollary}{\hspace{0pt}\bf Corollary}

\newtheorem{remark}{\hspace{0pt}\bf Remark}

\newtheorem{definition}{\hspace{0pt}\bf Definition}

\begin{document}

	
\title{
 Sampling of  Signals on Multigraphs: From Removable and Uniqueness sets to Spectral Proxies
	}


\author{Alejandro Parada-Mayorga\hspace{0.8mm}\orcidlink{0000-0003-4425-1816}
\thanks{A. Parada-Mayorga is with the Department of Electrical Engineering, University of Colorado, Denver, USA (e-mail: alejandro.paradamayorga@ucdenver.edu).}}


\markboth{IEEE Transactions on Signal Processing (SUBMITTED)}%
{Parada-Mayorga: Sampling of Signals on Multigraphs}
\maketitle





\begin{abstract}
In this paper, we study the sampling of bandlimited signals on multigraphs,
where several graphs with different shift operators are defined on the same
set of nodes. Since the shift operators do not commute in general, they
admit only a joint block diagonalization, and the frequencies of the
multigraph signal model are matrices rather than scalars. Relying on a
formal algebraic convolutional signal model, we introduce a notion of
frequency and bandwidth for multigraph signals that does not depend on any
individual layer, and we extend the notions of removable and uniqueness
sets to multigraphs. We prove that the optimality of a sampling set is
determined exclusively by the common Fourier basis and its invariant
subspaces, and not by any individual layer, although the individual layers
provide computable sufficient conditions that certify a sampling set as a
uniqueness set. Since identifying which layers provide reliable certificates
requires spectral information, we propose an aggregate operator that pools
the information of all the layers, is computable directly from the shift
operators, and certifies uniqueness without any assumption on the
individual layers. This operator leads to a greedy sampling algorithm that
does not require computing the joint block diagonalization. Numerical
experiments on multigraphs whose layers order the frequencies
inconsistently show that the proposed algorithm selects better-conditioned
sampling sets and attains lower reconstruction errors than single-layer
methods.
\end{abstract}

\IEEEpeerreviewmaketitle



%
%

\section{Introduction}
\label{sec_intro}

Sampling is a prevalent and essential problem in the processing of
information defined on irregular and graph-like structures. Whether the goal is to decide
where to place a limited number of sensors, which users to query in a
recommender system, or which regions to measure in a brain network, the
question is the same: on which subset of the nodes must a signal be
observed so that it can be recovered from those observations alone? In
graph signal processing (GSP)~\cite{ortega_gsp}, 
this problem is fairly well studied for bandlimited signals, and several
complementary perspectives are available, including the characterization
of uniqueness sets through removable sets~\cite{pesensonams1,pesenson2009}, 
sampling theorems for bandlimited graph signals~\cite{chen_samp,tsisvero_samp}, 
greedy selection based on spectral
proxies~\cite{ortega_proxies}, and random sampling
strategies~\cite{puy_samp}. 
These approaches share two ingredients: a notion of frequency, provided by
the eigendecomposition of a single shift operator, and an ordering of those
frequencies, which defines what it means for a signal to be bandlimited.

In many applications, however, the nodes are related to each other in more
than one way, and the natural domain of the signals is a \emph{multigraph}:
a collection of graphs -- or \emph{layers} -- that share the same node set,
each one with its own shift operator~\cite{msp_j,algnn_nc_j}. 
Signal processing on multigraphs is rigorously grounded in algebraic signal
processing~\cite{msp_j,algnn_nc_j}: 
the signals are defined once on the common node set, and the convolutional
filters are polynomials in the shift operators of the layers. Since these
operators do not commute in general, they cannot be jointly diagonalized,
and the best one can expect is a \emph{joint block diagonalization}, in
which the minimal invariant subspaces have dimension larger than one, the
frequencies are matrices rather than scalars, and each layer orders the
invariant subspaces in its own way. Consequently, neither of the two
ingredients of GSP sampling is available in its usual form, and, to the
best of our knowledge, the sampling problem for multigraph signals has
remained open. Existing approaches for structures with several graphs, such
as product graphs~\cite{samp_prodG_ortiz} 
and tensor models of multilayer networks~\cite{zhang2021mgsp,sardellitti2019multilayer}, 
do not cover this setting, since they define one signal per layer or on a
product of node sets, and they rely on a separable or otherwise prescribed
coupling between the layers. The basic questions of sampling theory must
therefore be revisited: What is a bandlimited signal on a multigraph? When
is a subset of the nodes a uniqueness set? Does the optimal sampling set
depend on which layer is used to measure the frequencies? And can such a
set be found without computing the joint block diagonalization?

In this paper, we answer these questions. We introduce a layer-free notion
of frequency, which associates to each invariant subspace the largest
singular value of its frequencies across all the layers, and we use it to
define the Paley-Wiener spaces of bandlimited multigraph signals. Following the approach of
Pesenson~\cite{pesensonams1,pesenson2009,fuhrpesenson} and its extension to graphon signals~\cite{gphon_samp}, 
we extend the notions of removable and uniqueness sets to multigraphs, and
we show that one informative layer suffices for uniqueness. The central
observation of the paper is that the exact optimality of a sampling set is
a property of the common Fourier basis of the multigraph, and not of any of
its layers. We introduce the \emph{spectral reach} of a sampling set, we
prove that it characterizes uniqueness sets exactly, and we show that, as a
consequence, any \emph{admissible} element of the filter algebra certifies
the same optimal sampling sets. Since admissibility cannot be verified
without the joint block diagonalization, and generically none of the layers
needs to be admissible, we introduce an aggregate operator that pools the
spectral information of all the layers and is computable directly from the
shift operators. We prove that it certifies uniqueness unconditionally,
and we use it to derive a greedy sampling algorithm that extends the
algorithm of~\cite{ortega_proxies} to multigraphs. In numerical
experiments on multigraphs where no layer is admissible, the proposed
algorithm attains the smallest reconstruction error in almost every
configuration, even when it is compared with the best single layer
selected \emph{a posteriori}.

This paper is organized as follows. In Section~\ref{sec_multigraph_sp}, we
introduce the signal model on multigraphs and the notion of bandlimited
multigraph signal. In Section~\ref{sec_removable_uniqueness}, we extend the
notions of removable and uniqueness sets to multigraphs. In
Section~\ref{sec_coupled_optimality}, we discuss the coupled optimality of
sampling sets in graphs and multigraphs, and we present the proposed
sampling algorithm. In Section~\ref{sec_experiments}, we present the
numerical experiments, and in Section~\ref{sec_conclusions}, we present
some conclusions. The proofs of all the results are presented in Appendix~\ref{app_proofs},
and the details of the generation of the multigraphs used in the
experiments in Appendix~\ref{app_multigraph_generation}; both appendices
are included in the supplementary material.





\section{Convolutional Signal Processing on Multigraphs}
\label{sec_multigraph_sp}




In our discussion, multigraphs are the domain on which we define the
information we want to process. To formally define multigraphs, we start by
stating some notation on graphs. We recall that a graph
$\ccalG=(\ccalV,\ccalE,w)$ is determined by a set of nodes $\ccalV$, a set of
edges $\ccalE$, and a weight function $w: \ccalE \to \mbR$. In our discussion,
the set $\ccalE$ is a set of unordered pairs, naturally associated with
undirected graphs. Then, the element $(i,j)\in\ccalE$ indicates that there is
an edge \textit{connecting} the nodes $i$ and $j$ in $\ccalV$, and it is
completely equivalent to the element $(j,i)$. On any element $(i,j)\in\ccalE$
there is an associated weight $w_{i,j}=w\{ (i,j) \}\in\mbR$ that provides a
measure of the strength of the connection between $i$ and $j$. The values
$w_{i,j}$ are stored in the so-called weight or adjacency matrix $\mathbf{W}$. In particular, $\mathbf{W}(i,j)=w_{i,j}$, where $(i,j)$ is the entry in row
$i$ and column $j$. Building on $\mathbf{W}$ it is possible to define on
$\ccalG$ other matrix operators, such as the combinatorial Laplacian, the
normalized Laplacian, and the random walk Laplacian, among others~\cite{ortega_gsp,dsp_graphsmoura},
which are referred to as \textit{shift operators} and represented with the
symbol $\mathbf{S}$. With this basic notation at hand, we are ready to
introduce the formal definition of a multigraph following~\cite{msp_j}.


\begin{definition}\label{def_multigraph}
A \emph{multigraph} $\ccalM = (\ccalV, \{\ccalE_r,w_{r}\}_{r=1}^{R})$
consists of a collection of graphs $\{ (\ccalV, \ccalE_{r},w_{r})
\}_{r=1}^{R}$ that share the same node set $\ccalV$. Additionally, if
$\mathbf{S}_{r}$ is the shift operator of the graph $(\ccalV,
\ccalE_{r},w_{r})$, then the collection $\{ \mathbf{S}_{r} \}_{r=1}^{R}$ is
called the collection of shift operators of $\ccalM$.
\end{definition}


\noindent Multigraphs thus provide a means of modeling several distinct types
of relationship among a common set of entities. It is important to emphasize
that the choice of shift operator is made independently for each graph. For
instance, one layer may be represented by an adjacency matrix and another by
a Laplacian. For our discussion, the particular choice of the shift operators
and their meaning is not central.


\subsection{Signals and Convolutional Filtering on Multigraphs}
Having fixed the domain, we now describe the data that lives on it. In graph
signal processing (GSP), a signal is a map from the vertex set into $\mbR$
and is identified with a vector in $\mbR^{|\ccalV|}$. Since a multigraph
shares a single node set across all of its layers, this notion extends to it
directly, as follows.


\begin{definition}\label{def_mult_signal}
Let $\ccalM = (\ccalV, \{\ccalE_r,w_{r}\}_{r=1}^{R})$ be a multigraph. A
\emph{signal} $\mathbf{x}$ on $\ccalM$ is a map $\mathbf{x} : \ccalV \to
\mbR$. To emphasize the domain of the signal we represent signals with the
symbol $(\ccalM,\mathbf{x})$.
\end{definition}


Once a labeling of the nodes is fixed, this map is identified with a vector
$\mathbf{x}\in \mbR^{|\ccalV|}$ carrying one entry per node, whose $i$th
component is the value of the signal at the node indexed by $i\in\ccalV$.

One more notion inherited intuitively from graphs and rigorously justified by
algebraic signal processing is that of a convolutional filter, which we
formally present in the following definition. Since the shift operators of a
multigraph do not commute in general, the filters must be described by
polynomials in \emph{noncommuting} variables. We recall that a
noncommutative polynomial in the variables $t_{1},\ldots,t_{R}$ is a finite
linear combination of \emph{words} $t_{w} = t_{r_{k}}t_{r_{k-1}}\cdots
t_{r_{1}}$ with $r_{j}\in\{1,\ldots,R\}$, where the length of the word is
$\vert w\vert = k$ and the empty word is identified with the unit.


\begin{definition}\label{def_mult_filter}
Let $\ccalM = (\ccalV, \{\ccalE_r,w_{r}\}_{r=1}^{R})$ be a multigraph with
shift operators $\{ \mathbf{S}_{r} \}_{r=1}^{R}$ and $N=\vert\ccalV\vert$,
and let $p$ be a noncommutative polynomial in $R$ variables. Then, we say
that $p\left( \mathbf{S}_{1}, \ldots, \mathbf{S}_{R} \right)\in\mbR^{N\times
N}$ is a \emph{convolutional filter} on $\ccalM$. Additionally, the action of
$p$ on a multigraph signal $(\ccalM,\mathbf{x})$ produces the signal
$(\ccalM,\mathbf{y})$, where
$
\mathbf{y}
       =
        p\left( \mathbf{S}_{1}, \mathbf{S}_{2}, \ldots, \mathbf{S}_{R} \right)
        \mathbf{x}
        .
$
\end{definition}


It is worth pointing out that although Definition~\ref{def_mult_filter}
presents filters on multigraphs as simple polynomials, their existence is
rigorously grounded in algebraic signal models based on algebras with
multiple generators. That is, this is not an ad hoc simplification of what a
filter is, but a particular instantiation of a general convolutional filter
in an algebraic signal model~\cite{algnn_nc_c}. In what follows, we denote by
$\ccalA(\mathbf{S}_{1},\ldots,\mathbf{S}_{R})$ the algebra of all
convolutional filters on $\ccalM$, i.e., the algebra generated by the shift
operators, and for a word $w=(r_{k},\ldots,r_{1})$ we write
$\mathbf{S}_{w}=\mathbf{S}_{r_{k}}\cdots\mathbf{S}_{r_{1}}$.


\subsection{Fourier Representations on Multigraphs}


\begin{figure*}
	\centering

\begin{tikzpicture}[
    >=Latex,
    font=\small,
    vertex/.style={circle,draw=aspnavy,fill=aspnavy!12,minimum size=5.5mm,inner sep=0pt,thick},
    edgeA/.style={aspnavy,line width=1pt},
    edgeB/.style={aspgold!65!black,line width=1pt,dashed},
]

\coordinate (leftanchor) at (0,0);

\def\cell{0.38}
\pgfmathsetmacro{\matside}{7*\cell}
\pgfmathsetmacro{\bA}{2*\cell}
\pgfmathsetmacro{\bB}{5*\cell}
\def\gapone{0.6}
\def\gaptwo{0.9}
\def\vecwidth{0.45}

\coordinate (S1org) at ($(leftanchor)+(9mm,1.47)$);
\coordinate (S2org) at ($(S1org)+(\matside+\gapone,0)$);
\coordinate (Xorg)  at ($(S1org)+(2*\matside+\gapone+\gaptwo,0)$);

\draw[aspgray,dashed,thick,rounded corners=1pt] (S1org) rectangle ++(\matside,-\matside);
\coordinate (b1cur) at (S1org);
\foreach \u/\col/\txt [count=\i] in {2/blockA/aspnavy,3/blockB/aspnavy,2/blockC/white} {
  \pgfmathsetmacro{\sz}{\u*\cell}
  \draw[fill=\col,draw=aspnavy,thick] (b1cur) rectangle ++(\sz,-\sz)
      node[midway,font=\normalsize,color=\txt] {$\mathbf{\Xi}_{\i}^{(1)}$};
  \path (b1cur) ++(\sz,-\sz) coordinate (b1cur);
}
\coordinate (S1top) at ($(S1org)+(0.5*\matside,0)$);

\draw[aspgray,dashed,thick,rounded corners=1pt] (S2org) rectangle ++(\matside,-\matside);
\coordinate (b2cur) at (S2org);
\foreach \u/\col/\txt [count=\i] in {2/blockA/aspnavy,3/blockB/aspnavy,2/blockC/white} {
  \pgfmathsetmacro{\sz}{\u*\cell}
  \draw[fill=\col,draw=aspnavy,thick] (b2cur) rectangle ++(\sz,-\sz)
      node[midway,font=\normalsize,color=\txt] {$\mathbf{\Xi}_{\i}^{(2)}$};
  \path (b2cur) ++(\sz,-\sz) coordinate (b2cur);
}
\coordinate (S2top) at ($(S2org)+(0.5*\matside,0)$);

\coordinate (xcur) at (Xorg);
\foreach \u/\col/\txt [count=\i] in {2/blockA/aspnavy,3/blockB/aspnavy,2/blockC/white} {
  \pgfmathsetmacro{\sz}{\u*\cell}
  \draw[fill=\col,draw=aspnavy,thick] (xcur) rectangle ++(\vecwidth,-\sz);
  \path (xcur) ++(0,-\sz) coordinate (xcur);
}
\coordinate (Xtop) at ($(Xorg)+(0.5*\vecwidth,0)$);
\node[above=2mm of Xtop,font=\normalsize] {$\widehat{\mathbf{x}}=\mathbf{U}^\top\mathbf{x}$};

\coordinate (guideXend) at ($(Xorg)+(\vecwidth,0)$);
\draw[aspgray,dotted,thick] ($(S1org)+(0,-\bA)$) -- ($(guideXend)+(0,-\bA)$);
\draw[aspgray,dotted,thick] ($(S1org)+(0,-\bB)$) -- ($(guideXend)+(0,-\bB)$);

\coordinate (braceR) at ($(Xorg)+(\vecwidth,0)$);
\draw[decorate,decoration={brace,amplitude=4pt,raise=1mm},aspnavy]
   (braceR) -- ($(braceR)+(0,-\bA)$)
   node[midway,right=3mm,font=\small,align=left]{$\in\mathbb{R}^{q_{1}}$};
\draw[decorate,decoration={brace,amplitude=4pt,raise=1mm},aspnavy]
   ($(braceR)+(0,-\bA)$) -- ($(braceR)+(0,-\bB)$)
   node[midway,right=3mm,font=\small,align=left]{$\in\mathbb{R}^{q_2}$};
\draw[decorate,decoration={brace,amplitude=4pt,raise=1mm},aspnavy]
   ($(braceR)+(0,-\bB)$) -- ($(braceR)+(0,-\matside)$)
   node[midway,right=3mm,font=\small,align=left]{$\in\mathbb{R}^{q_3}$};


\coordinate (specbottom) at ($(S1org)+(1.5*\matside,-1*\matside)$);
\coordinate (spectop) at ($(S1org)+(1.2*\matside,0)$);
\node[above=2mm of spectop,font=\small,align=center] (eqn)
  {$\mathbf{S}_r=\mathbf{U}\,\mathrm{diag}\left(\mathbf{\Xi}_1^{(r)},\mathbf{\Xi}_2^{(r)},\mathbf{\Xi}_3^{(r)}\right)\,\mathbf{U}^{\top}$};

   \node[below=3mm of specbottom,font=\small,align=center] (eqn1)
   {$\text{dim}\left(\mathbf{\Xi}_{\ell}^{(r)}\right)=q_{\ell}\times q_{\ell}$ for every $r$};


\node[below=3mm of eqn] (row1lab) {};

\coordinate (connA) at ($(row1lab)+(1cm,0)$);
\coordinate (connB) at ($(connA)+(1cm,0)$);


\coordinate (Porg) at ($(S1org)+(3.35*\matside,0*\matside)$);

\draw[aspgray,dashed,thick,rounded corners=1pt] (Porg) rectangle ++(\matside,-\matside);
\coordinate (bpcur) at (Porg);
\foreach \u/\col/\txt [count=\i] in {2/blockA/aspnavy,3/blockB/aspnavy,2/blockC/white} {
  \pgfmathsetmacro{\sz}{\u*\cell}
  \draw[fill=\col,draw=aspnavy,thick] (bpcur) rectangle ++(\sz,-\sz)
      node[midway,font=\small,color=\txt] {$\mathbf{P}_{\i}$};
  \path (bpcur) ++(\sz,-\sz) coordinate (bpcur);
}
\coordinate (Ptop) at ($(Porg)+(0.5*\matside,0)$);
\node[above=2mm of Ptop,font=\small] {$\mathrm{diag}\big(\mathbf{P}_1,\mathbf{P}_2,\mathbf{P}_3\big)$};

\coordinate (PbraceR) at ($(Porg)+(1*\matside,0)$);
\draw[decorate,decoration={brace,amplitude=4pt,raise=1mm},aspnavy]
   (PbraceR) -- ($(PbraceR)+(0,-\bA)$)
   node[midway,right=3mm,font=\small,align=left]{$\mathbf{P}_1=p\left(\mathbf{\Xi}_1^{(1)},\mathbf{\Xi}_1^{(2)}\right)$};
\draw[decorate,decoration={brace,amplitude=4pt,raise=1mm},aspnavy]
   ($(PbraceR)+(0,-\bA)$) -- ($(PbraceR)+(0,-\bB)$)
   node[midway,right=3mm,font=\small,align=left]{$\mathbf{P}_2=p\left(\mathbf{\Xi}_2^{(1)},\mathbf{\Xi}_2^{(2)}\right)$};
\draw[decorate,decoration={brace,amplitude=4pt,raise=1mm},aspnavy]
   ($(PbraceR)+(0,-\bB)$) -- ($(PbraceR)+(0,-\matside)$)
   node[midway,right=3mm,font=\small,align=left]{$\mathbf{P}_3=p\left(\mathbf{\Xi}_3^{(1)},\mathbf{\Xi}_3^{(2)}\right)$};

\coordinate (X2org) at ($(PbraceR)+(4,0)$);
\coordinate (xcur2) at (X2org);
\foreach \u/\col/\txt [count=\i] in {2/blockA/aspnavy,3/blockB/aspnavy,2/blockC/white} {
  \pgfmathsetmacro{\sz}{\u*\cell}
  \draw[fill=\col,draw=aspnavy,thick] (xcur2) rectangle ++(\vecwidth,-\sz);
  \path (xcur2) ++(0,-\sz) coordinate (xcur2);
}
\coordinate (X2top) at ($(X2org)+(0.5*\vecwidth,0)$);
\node[above=2mm of X2top,font=\small] {$\widehat{\mathbf{x}}=\mathbf{U}^\top\mathbf{x}$};

\node[font=\Large,aspnavy] (times2) at ($(X2org)+(-0.5,-0.5*\matside)$) {$\times$};

\coordinate (eqsign) at ($(X2org)+(\vecwidth+1*\gapone,-0.5*\matside)$);
\node[font=\Large,aspnavy] at (eqsign) {$=$};

\coordinate (Y2org) at ($(X2org)+(\vecwidth+2*\gapone,0)$);
\coordinate (ycur2) at (Y2org);
\foreach \u/\col/\txt [count=\i] in {2/blockA/aspnavy,3/blockB/aspnavy,2/blockC/white} {
  \pgfmathsetmacro{\sz}{\u*\cell}
  \draw[fill=\col,draw=aspnavy,thick] (ycur2) rectangle ++(\vecwidth,-\sz);
  \path (ycur2) ++(0,-\sz) coordinate (ycur2);
}
\coordinate (Y2top) at ($(Y2org)+(0.5*\vecwidth,0)$);
\node[above=2mm of Y2top,font=\small] {$\widehat{\mathbf{y}}$};

\coordinate (row2bottom) at ($(Porg)+(1.7*\matside,-\matside)$);
\node[below=3mm of row2bottom,font=\small,align=center] (eqn2)
  {$p(\mathbf{S}_1,\mathbf{S}_2)=\mathbf{U}\,\mathrm{diag}\big(\mathbf{P}_1,\mathbf{P}_2,\mathbf{P}_3\big)\,\mathbf{U}^{\top}
   \Rightarrow   \mathrm{diag}\big(\mathbf{P}_1,\mathbf{P}_2,\mathbf{P}_3\big)\,\widehat{\mathbf{x}}=\widehat{\mathbf{y}}$};

\end{tikzpicture}
	\caption{Fourier representations on a multigraph. The shift operators
	$\{\mathbf{S}_{r}\}_{r=1}^{R}$ are jointly block diagonalized by a common
	orthonormal basis $\mathbf{U}$. The components of
	$\widehat{\mathbf{x}}=\mathbf{U}^{\top}\mathbf{x}$ are grouped by the
	blocks of the partition, and convolutional filters act blockwise on the
	spectral representation.}
	\label{fig_multi_fourier}
\end{figure*}
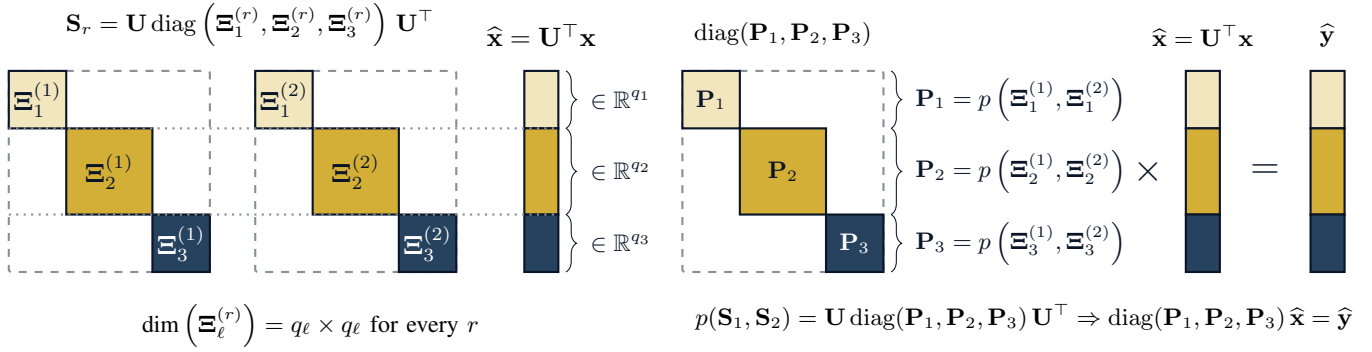


%
In any convolutional signal processing model, the notion of a spectral or
Fourier representation is rooted in the algebraic notion of
irreducibility~\cite{algnn_nc_j,parada_algnn,repthybigbook}, which captures the essential notion of
\textit{invariance} with respect to the action of the filters in the signal
model. Then, those minimal invariant subspaces are used to build a formal
decomposition of the space of signals that, at the same time, translates the
filters into local operators over each invariant subspace.

In GSP this specializes to the spectral decomposition of the shift operator
in terms of its eigenvectors, since the irreducible subrepresentations in
this case are one-dimensional~\cite{algSP0,parada_algnn}. This translates the representation of a
convolutional graph filter into a polynomial acting on the eigenvalues of the
shift operator~\cite{dsp_graphsmoura,gsp_graphfilters,parada_algnn}.

In a multigraph, the arbitrary nature of the shift operators $\{
\mathbf{S}_{r} \}_{r=1}^{R}$ typically leads to invariant subspaces of
dimension larger than one. This translates into a concrete scenario where
the $\{ \mathbf{S}_{r} \}_{r=1}^{R}$ are jointly \emph{block}
diagonalizable, instead of jointly diagonalizable, which is the case when all
the shift operators commute.

To properly establish the notion of Fourier transform on multigraphs, we
start by recalling the basic concept of block diagonalizable matrices. For an
integer $Q \in \mathbb{N}$, a \emph{partition} $\tau_Q = (q_1, \dots, q_L)$
is a tuple of positive integers with $\sum_{\ell=1}^{L} q_\ell = Q$. Then, a
collection of matrices $\{\mathbf{S}_r\}_{r=1}^{R}$, with $\mathbf{S}_r \in
\mbR^{N \times N}$, admits a \emph{joint block diagonalization} if there
exist a partition $\tau_N = (q_1, \dots, q_L)$ and an orthonormal 
$\mathbf{U}\in\mbR^{N\times N}$ such that
\begin{equation}\label{eq_jbd}
\mathbf{S}_r =
     \mathbf{U}
          \operatorname{diag}\left( 
                                \mathbf{\Xi}_1^{(r)}, \dots, \mathbf{\Xi}_L^{(r)}
                                 \right) 
     \mathbf{U}^{\top},
  \quad\mathbf{\Xi}_\ell^{(r)} \in \mbR^{q_\ell \times q_\ell},
\end{equation}
for every $r = 1, \dots, R$. Throughout, we assume that the partition is
\emph{maximal}, i.e., that the blocks cannot be refined further by any
orthonormal change of basis. This guarantees that the subspaces spanned by
the columns of $\mathbf{U}$ associated with each block are minimal invariant
subspaces of the signal model. We denote by
$\mathbf{U}_{\ell}\in\mbR^{N\times q_{\ell}}$ the submatrix of $\mathbf{U}$
whose columns are associated with the $\ell$th block, and by
$\widehat{\mathbf{x}}_{\ell}=\mathbf{U}_{\ell}^{\top}\mathbf{x}\in\mbR^{q_{\ell}}$
the corresponding group of coefficients -- see Fig.~\ref{fig_multi_fourier}.

Then, in the case of a multigraph, the existence of a joint block
diagonalization of the shift operators guarantees that there is an
orthonormal basis $\mathbf{U}\in\mbR^{N\times N}$ that can be used for a
\textit{decomposition} of any multigraph signal $(\ccalM,\mathbf{x})$.
Additionally, such an orthonormal basis encapsulates the invariances and
subspaces dictated by the sizes of the blocks $\mathbf{\Xi}_{\ell}^{(r)}\in
\mbR^{q_\ell \times q_\ell}$. We leverage these concepts to formally
introduce the notion of the multigraph Fourier transform in the following
definition.


\begin{definition}\label{def_mft}
Let $\ccalM$ be a multigraph with shift operators $\{ \mathbf{S}_{r}
\}_{r=1}^{R}$ satisfying~\eqref{eq_jbd} with
$\mathbf{U}\mathbf{U}^\top =\mathbf{I}=\mathbf{U}^\top\mathbf{U}$. Then, we
say that $(\widehat{\ccalM},\widehat{\mathbf{x}})$ is the \emph{multigraph
Fourier transform} of the signal $(\ccalM,\mathbf{x})$, with
$\widehat{\mathbf{x}}=\mathbf{U}^\top \mathbf{x}$ and $\widehat{\ccalM}$
denoting the collection of blocks $\{\mathbf{\Xi}_{\ell}^{(r)}\}_{\ell,r}$. We refer to the
$\mathbf{\Xi}_\ell^{(r)}$ as the \emph{frequencies}. The inverse multigraph
Fourier transform is determined by the action of $\mathbf{U}$, i.e.,
$\mathbf{x}=\mathbf{U}\widehat{\mathbf{x}}$.
\end{definition}


\noindent Then, we can see that the dimensions of the invariant subspaces of
the signal model on $\ccalM$ are $\{ q_{\ell} \}_{\ell=1}^{L}$. Notice that
joint diagonalization -- i.e., $q_{\ell}=1$ for all $\ell$ -- is guaranteed
when $\mathbf{S}_{i}\mathbf{S}_{j}=\mathbf{S}_{j}\mathbf{S}_{i}$ for all
$i,j$. Notice also that when a joint block diagonalization exists,
polynomials in the $\{ \mathbf{S}_{r}\}_{r=1}^{R}$ translate into polynomials
in $\mathbf{\Xi}_{\ell}^{(r)}$. That is, for any convolutional filter
$p(\mathbf{S}_{1},\ldots, \mathbf{S}_{R})$ acting on the multigraph signal
$(\ccalM,\mathbf{x})$ it follows that
\begin{multline}\label{eq_filter_node_spectral}
p\left( \mathbf{S}_{1},\ldots, \mathbf{S}_{R} \right)
          \mathbf{x}
          =
          \\
          \mathbf{U}
          \operatorname{diag}
             \left(
                    p\left(
                         \mathbf{\Xi}_{1}^{(1)},
                         \ldots,
                         \mathbf{\Xi}_{1}^{(R)}
                     \right)
                     ,
                     \ldots
                     ,
                     p\left(
                         \mathbf{\Xi}_{L}^{(1)},
                         \ldots,
                         \mathbf{\Xi}_{L}^{(R)}
                     \right)
              \right)
              \widehat{\mathbf{x}}.
\end{multline}
That is, the action of a filter on the node domain can be obtained as the
inverse multigraph Fourier transform of the spectrally filtered version of
the multigraph signal.

Notice that in the frequency domain the filter can be thought of as defined
pointwise on the frequencies and acting pointwise on the invariant subspaces
associated with the frequencies. Since the polynomial $p$ is noncommutative
and the blocks $\mathbf{\Xi}_{\ell}^{(r)}$ of a given index $\ell$ do not
commute in general, the order of the letters in each word of $p$ is
preserved when evaluating
$p\left( \mathbf{\Xi}_{\ell}^{(1)},\ldots,\mathbf{\Xi}_{\ell}^{(R)}\right)$.


\begin{remark}
\label{rem_freq_ordering}\normalfont
On multigraphs, the nature of the frequencies is multivariate: each invariant
subspace $\ell$ carries an $R$-tuple of matrix frequencies
$\left( \mathbf{\Xi}_{\ell}^{(1)},\ldots,\mathbf{\Xi}_{\ell}^{(R)}\right)$, so that when
thinking of something like a \textit{frequency axis} one should associate one
axis to each of the shift operators of the multigraph. Then, there will be $R$ frequency axes, and
this is completely independent of how the invariances take place, i.e., of
the dimensions of the blocks $\mathbf{\Xi}_{\ell}^{(r)}$. For instance, if
the shift operators commute, then $\mathbf{\Xi}_{\ell}^{(r)}\in\mbR^{1\times
1}$ and the spectral representation of a filter is determined by a
\textit{scalar} multivariate polynomial. However, if the dimension of at
least one block is larger than one, the spectral representation of the filter
is determined by a multivariate \textit{matrix} polynomial. That is, the
number of independent variables in those polynomials is the same, but their
nature differs.

For the developments in this paper we require a total ordering of the
blocks. Since the shift operators associated with each layer of the multigraph induce, in general, \emph{different} orderings --
requiring $\sigma_{\max}(\mathbf{\Xi}_{\ell}^{(r)})$ to be simultaneously
nondecreasing in $\ell$ for every $r$ is generically not satisfiable -- we
adopt a canonical, layer-free convention. To each block we associate the
\emph{frequency label}
\begin{equation}\label{eq_freq_label}
\mu_{\ell}
   =
   \max_{r=1,\ldots,R}
       \sigma_{\max}\!\left(\mathbf{\Xi}_{\ell}^{(r)}\right)
       ,
\end{equation}
and we order the blocks so that
\begin{equation}\label{eq_freq_ordering}
\mu_{1}
      \leq
         \mu_{2}
               \leq
                  \ldots
                  \leq
                   \mu_{L}
                   .
\end{equation}
This convention resembles the typical convention in GSP of ordering the
frequencies associated with the graph Fourier transform (GFT), while
aggregating the $R$ frequency axes into a single worst-case axis. When the
shift operators commute and $R=1$, \eqref{eq_freq_ordering} reduces exactly
to the usual GSP ordering by $\vert\lambda_{\ell}\vert$.

\end{remark}



\subsection{Bandwidth of a Multigraph Signal}

Once the ordering of the frequencies in~\eqref{eq_freq_ordering} is fixed,
it is possible to establish the notion of bandlimitedness of a multigraph
signal, following a natural extension of the same concept from GSP. To do so
formally, it is essential to emphasize -- as in
Fig.~\ref{fig_multi_fourier} -- that while $\widehat{\mathbf{x}}$ has
$N=\vert \ccalV\vert$ components, the association of those components with
the frequency domain is determined by the blocks in the joint block
diagonalization of the shift operators. For the particular case of
Fig.~\ref{fig_multi_fourier}, the first Fourier components of
$\widehat{\mathbf{x}}$ are associated with the blocks
$\mathbf{\Xi}_{1}^{(1)}$ and $\mathbf{\Xi}_{1}^{(2)}$, precisely because the dimension of the first invariant subspace is $q_1$. In this context, we refer
to $\widehat{\mathbf{x}}_{1}=[\widehat{\mathbf{x}}(1), \widehat{\mathbf{x}}(2),\ldots,\widehat{\mathbf{x}}(q_1)]^{\top}$ as the Fourier components of
$\widehat{\mathbf{x}}$ defined on
$\left(\mathbf{\Xi}_{1}^{(1)},\mathbf{\Xi}_{1}^{(2)}\right)$. With this
context at hand, we introduce the following definition.


\begin{definition}\label{def_bandlimited_signal}
Let $(\ccalM,\mathbf{x})$ be a multigraph signal on $\ccalM$ with shift
operators $\{ \mathbf{S}_{r} \}_{r=1}^{R}$ satisfying~\eqref{eq_jbd}, and
let $\omega\geq 0$. We say that $(\ccalM,\mathbf{x})$ is
\emph{$\omega$-bandlimited} if
\begin{equation}\label{eq_def_bandlimited}
\widehat{\mathbf{x}}_{\ell}=\mathbf{0}
\quad
\text{for all }\ell\text{ such that }\mu_{\ell}>\omega,
\end{equation}
where $(\widehat{\ccalM},\widehat{\mathbf{x}})$ is the multigraph Fourier
transform of $(\ccalM,\mathbf{x})$ and $\mu_{\ell}$ is the frequency label
in~\eqref{eq_freq_label}. Equivalently, $\widehat{\mathbf{x}}_{\ell}=
\mathbf{0}$ whenever
$\sigma_{\max}(\mathbf{\Xi}_{\ell}^{(r)})>\omega$ for some
$r\in\{1,\ldots,R\}$. The set of all $\omega$-bandlimited signals on $\ccalM$
is denoted by $\mathcal{PW}_{\omega}(\ccalM)$.
\end{definition}


\noindent Notice from Definition~\ref{def_bandlimited_signal} that the
multivariate nature of the frequency components is organically covered by the
definition: if one thinks in terms of the $R$ frequency axes in the
representation of the signal, the bandwidth is established as the same value
$\omega$ on all axes, and a group of Fourier components survives only if it
is below the bandwidth on \emph{every} axis. Additionally, denoting by
$\ell(\omega)=\max\{\ell : \mu_{\ell}\leq\omega\}$ (with $\ell(\omega)=0$ if
$\mu_{1}>\omega$) and by $\ccalR(\omega)$ the set of column indices of
$\mathbf{U}$ associated with the first $\ell(\omega)$ blocks, we have the
compact characterization
\begin{equation}\label{eq_pw_as_range}
\mathcal{PW}_{\omega}(\ccalM)
   =
   \operatorname{range}
      \left(
          \mathbf{U}_{\ccalR(\omega)}
      \right)
   =
   \bigoplus_{\ell=1}^{\ell(\omega)}
       \operatorname{range}
          \left(
              \mathbf{U}_{\ell}
          \right)
      ,
\end{equation}
which shows that $\mathcal{PW}_{\omega}(\ccalM)$ is a subspace of
$\mbR^{N}$ of dimension $\sum_{\ell\leq\ell(\omega)}q_{\ell}$.

Notice also that when talking about the bandwidth in terms of the number of
components of $\widehat{\mathbf{x}}$, it only makes sense to do so taking
into account the number and dimensions of the blocks
$\mathbf{\Xi}_{\ell}^{(r)}$. That is, while the value of the bandwidth
$\omega$ can be arbitrarily stated on the spectral axes, the number of
components associated with each bandwidth has a limited number of
possibilities, i.e., the dimension of $\mathcal{PW}_{\omega}(\ccalM)$ can only take
the $L+1$ values $\{0,\, q_{1},\, q_{1}+q_{2},\ldots,\, N\}$. For instance,
in the case of Fig.~\ref{fig_multi_fourier}, and assuming all the labels
$\mu_{\ell}$ are different, any multigraph signal can have only four
possible bandwidth values when looking at the components of the vector
$\widehat{\mathbf{x}}$ directly. This represents a significant contrast with
the classical GSP scenario, where on a graph with $N$ nodes the dimension of
the bandlimited space can take any value in $\{0,1,\ldots, N\}$. As we will
see, this \emph{quantization} of the admissible bandlimited spaces is one of
the structural traits that distinguishes sampling on multigraphs from
sampling on graphs.

Before closing this section, we emphasize a structural property of
$\mathcal{PW}_{\omega}(\ccalM)$ that will be essential for subsequent discussions. Since bandlimitedness in Definition~\ref{def_bandlimited_signal} is supported
on \emph{whole} invariant subspaces, and every shift operator preserves every
invariant subspace, the bandlimited spaces are invariant not only under each
$\mathbf{S}_{r}$ individually but under the entire algebra of convolutional
filters.


\begin{lemma}\label{lemma_pw_invariance}
Let $\ccalM$ be a multigraph with shift operators $\{ \mathbf{S}_{r}
\}_{r=1}^{R}$ satisfying~\eqref{eq_jbd}. Then, for every $\omega\geq 0$ and
every convolutional filter
$p(\mathbf{S}_{1},\ldots,\mathbf{S}_{R})\in\ccalA(\mathbf{S}_{1},\ldots,
\mathbf{S}_{R})$, it holds that
\begin{equation}
p(\mathbf{S}_{1},\ldots,\mathbf{S}_{R})\,
\mathcal{PW}_{\omega}(\ccalM)
\subseteq
\mathcal{PW}_{\omega}(\ccalM)
.
\end{equation}
In particular, $\mathbf{S}_{w}\,\mathcal{PW}_{\omega}(\ccalM)\subseteq
\mathcal{PW}_{\omega}(\ccalM)$ for every word $w$.
\end{lemma}

\begin{proof}
    See Appendix~\ref{proof_lemma_pw_invariance}.
\end{proof}

We close this section by showing formally how the bandwidth of a multigraph
signal limits the norm of its diffusions through the shift operators. This
result, together with its corollary for words, will be fundamental to
establishing the sampling results of the following sections.


\begin{theorem}\label{thm_bernstein_ineq}
Let $(\ccalM,\mathbf{x})$ be a multigraph signal on the multigraph $\ccalM$
with shift operators $\{ \mathbf{S}_{r} \}_{r=1}^{R}$. If
$(\ccalM,\mathbf{x})$ is $\omega$-bandlimited, it follows that
\begin{equation}\label{eq_thm_bernstein_ineq_1}
\left\Vert
     \mathbf{S}_{r}
     \mathbf{x}
\right\Vert_{2}
              \leq
                  \omega
                  \left\Vert
                      \mathbf{x}
                  \right\Vert_{2}
                  \quad
                  \forall ~r=1,\ldots, R
                  .
\end{equation}
\end{theorem}

\begin{proof}
    See Appendix~\ref{proof_thm_bernstein_ineq}.
\end{proof}

The invariance stated in Lemma~\ref{lemma_pw_invariance} allows us to iterate
Theorem~\ref{thm_bernstein_ineq} along arbitrary words of shift operators.
The resulting inequality is the multigraph counterpart of the powers of the
variation operator used in~\cite{ortega_proxies}, and it is the property
that licenses the higher-order sampling machinery of
Section~\ref{sec_coupled_optimality}.


\begin{corollary}\label{cor_word_bernstein}
Let $(\ccalM,\mathbf{x})$ be an $\omega$-bandlimited multigraph signal on
$\ccalM$. Then, for every word $w=(r_{k},\ldots,r_{1})$ of length $\vert
w\vert = k$,
\begin{equation}\label{eq_cor_word_bernstein}
\left\Vert
     \mathbf{S}_{w}
     \mathbf{x}
\right\Vert_{2}
   =
\left\Vert
     \mathbf{S}_{r_{k}}
     \cdots
     \mathbf{S}_{r_{1}}
     \mathbf{x}
\right\Vert_{2}
              \leq
                  \omega^{k}
                  \left\Vert
                      \mathbf{x}
                  \right\Vert_{2}
                  .
\end{equation}
In particular, $\Vert \mathbf{S}_{r}^{k}\mathbf{x}\Vert_{2}\leq
\omega^{k}\Vert\mathbf{x}\Vert_{2}$ for every $r$ and every
$k\in\mathbb{N}$.
\end{corollary}

\begin{proof}
    See Appendix~\ref{proof_cor_word_bernstein}.
\end{proof}




\section{Removable and Uniqueness Sets in Multigraphs}
\label{sec_removable_uniqueness}
The concepts of removable and uniqueness sets were introduced by Pesenson for
the analysis of bandlimited functions on graphs, manifolds, and Banach
spaces~\cite{pesensonams1,pesenson2009,pesenson2019}. Their definition, although simple, provides a powerful and
elegant tool for the characterization of sampling sets. Building on the
understanding of this tool, the notion was later extended
in~\cite{gphon_samp} for the study of signals on graphons. We now extend this concept to signals
on multigraphs, as a tool to characterize sampling sets in a way that is
compatible with the characterization of bandlimited signals given in
Definition~\ref{def_bandlimited_signal}. For a subset
$\ccalS\subseteq\ccalV$, we recall that
$
L_{2}(\ccalS)
   =
   \left\lbrace
       \mathbf{x}\in\mbR^{N}
       \; : \;
       \mathbf{x}\vert_{\ccalS^{c}}=\mathbf{0}
   \right\rbrace
$
denotes the set of all signals with support contained in $\ccalS$, and that
for a matrix $\mathbf{A}\in\mbR^{N\times N}$ we write
$[\mathbf{A}]_{\ccalS,\ccalS}$ for the
principal submatrix of $\mathbf{A}$ with rows and columns indexed by
$\ccalS$.


\begin{definition}\label{def_remov_set}
Let $\ccalM = (\ccalV, \{\ccalE_r,w_{r}\}_{r=1}^{R})$ be a multigraph with
shift operators $\{ \mathbf{S}_{r} \}_{r=1}^{R}$, and let
$\ccalS\subset\ccalV$ be nonempty. The \emph{removable constant of $\ccalS$
on the $\mathbf{S}_{r}$-axis} is
\begin{equation}\label{eq_def_remov_set_1}
\Lambda_{r,\ccalS}
   =
   \inf_{\mathbf{x}\in L_{2}(\ccalS),\ \mathbf{x}\neq\mathbf{0}}
      \frac{
            \left\Vert
                 \mathbf{S}_{r}\mathbf{x}
            \right\Vert_{2}
           }
          {
           \left\Vert
                 \mathbf{x}
           \right\Vert_{2}
          }
          ,
\end{equation}
so that $\Vert \mathbf{S}_{r}\mathbf{x}\Vert_{2}\geq
\Lambda_{r,\ccalS}\Vert\mathbf{x}\Vert_{2}$ for all $\mathbf{x}\in
L_{2}(\ccalS)$, and $\Lambda_{r,\ccalS}$ is the largest constant with this
property. We say that $\ccalS$ is \emph{removable on the
$\mathbf{S}_{r}$-axis} if $\Lambda_{r,\ccalS}>0$. Additionally, we define the
aggregated constants
\begin{equation}\label{eq_def_lambda_min_max}
\Lambda_{\ccalS}^{\min}
     =
      \min_{r=1,\ldots,R}
           \Lambda_{r,\ccalS}
           ,
           \qquad
\Lambda_{\ccalS}^{\max}
     =
      \max_{r=1,\ldots,R}
           \Lambda_{r,\ccalS}
           ,
\end{equation}
and we say that $\ccalS$ is \emph{removable} if
$\Lambda_{\ccalS}^{\max}>0$, i.e., if it is removable on at least one axis.
When the axis under consideration is clear from the context, we refer to
$\ccalS$ as a $\Lambda$-removable set with
$\Lambda\in\{\Lambda_{\ccalS}^{\min},\Lambda_{\ccalS}^{\max}\}$ as
appropriate.
\end{definition}


\noindent From Definition~\ref{def_remov_set} it is worth emphasizing a couple of points. First, since $L_{2}(\ccalS)$ is a finite dimensional subspace, the infimum
in~\eqref{eq_def_remov_set_1} is attained, and it admits an exact
computational characterization as an eigenvalue problem. This is the content
of the following lemma, which also settles exactly when the removable
constants are strictly positive.


\begin{lemma}
\label{lemma_lambda_eig}
Let $\ccalS\subset\ccalV$ be nonempty. Then
\begin{equation}\label{eq_lambda_eig}
\Lambda_{r,\ccalS}
   =
   \sqrt{
       \lambda_{\min}
          \!\left(
              \left[
                  \mathbf{S}_{r}^{\top}\mathbf{S}_{r}
              \right]_{\ccalS,\ccalS}
           \right)
        }
   =
   \sigma_{\min}
      \!\left(
          \mathbf{S}_{r}\,\mathbf{I}_{\ccalV\ccalS}
       \right)
   ,
\end{equation}
where $\mathbf{I}_{\ccalV\ccalS}$ denotes the submatrix of the identity whose
columns are indexed by $\ccalS$. Moreover, $\Lambda_{r,\ccalS}>0$ if and only
if
\begin{equation}\label{eq_kernel_condition}
\ker(\mathbf{S}_{r})
   \cap
   L_{2}(\ccalS)
   =
   \{\mathbf{0}\}
   ,
\end{equation}
i.e., if and only if the columns of $\mathbf{S}_{r}$ indexed by $\ccalS$ are
linearly independent.
\end{lemma}

\begin{proof}
    See Appendix~\ref{proof_lemma_lambda_eig}.
\end{proof}


\begin{remark}\label{rem_degenerate_sets}\normalfont
Condition~\eqref{eq_kernel_condition} shows that, contrary to what a first
reading of~\eqref{eq_def_remov_set_1} may suggest, \emph{not} every subset
of $\ccalV$ is removable on every axis. A natural example arises when the
layer $r$ is endowed with a combinatorial Laplacian and $\ccalS$ contains an
entire connected component of $(\ccalV,\ccalE_{r},w_{r})$. Then, the indicator
vector of that component belongs to $L_{2}(\ccalS)\cap\ker(\mathbf{S}_{r})$,
so $\Lambda_{r,\ccalS}=0$. Under the aggregated notion of
Definition~\ref{def_remov_set}, however, such a set may still be removable,
provided that at least one other layer satisfies the kernel
condition~\eqref{eq_kernel_condition}. This is a clear indication of the fact that, even when the sampling problem on one layer has degeneracies in terms of the constants $\Lambda_{r,\mathcal{S}}$, it can still be well defined on the multilayer structure. 
\end{remark}


Second, the constants $\Lambda_{r,\ccalS}$ can be directly compared with the
bandwidth of a signal to establish whether a sampling set allows the unique
identification of bandlimited signals. This is the content of the following
theorem. Notice that, in view of the Bernstein inequality of
Theorem~\ref{thm_bernstein_ineq} -- which controls the diffusion of a
bandlimited signal on \emph{every} axis simultaneously -- it suffices that
the removability inequality hold on a \emph{single} axis for uniqueness to
follow. Accordingly, the sharp sufficient condition is stated in terms of
$\Lambda^{\max}$ rather than $\Lambda^{\min}$.


\begin{theorem}\label{thm_unique_set}
Let $\ccalM = (\ccalV, \{\ccalE_r,w_{r}\}_{r=1}^{R})$ be a multigraph and
let $\ccalS\subset\ccalV$ be such that $\ccalS^{c}$ is removable, with
aggregated constant $\Lambda_{\ccalS^{c}}^{\max}$ as in
Definition~\ref{def_remov_set}. Let $(\ccalM,\mathbf{x})$ and
$(\ccalM,\mathbf{y})$ be $\omega$-bandlimited signals on $\ccalM$. If
\begin{equation}\label{eq_thm_unique_condition}
\omega<\Lambda_{\ccalS^c}^{\max}
\end{equation}
and $\mathbf{x}\vert_{\ccalS}=\mathbf{y}\vert_{\ccalS}$, then
$\mathbf{x}=\mathbf{y}$, and we refer to $\ccalS$ as a \emph{uniqueness set}
for $\omega$-bandlimited signals.
\end{theorem}

\begin{proof}
    See Appendix~\ref{proof_thm_unique_set}.
\end{proof}

\noindent Theorem~\ref{thm_unique_set} establishes a characterization of a
sampling set as a uniqueness set for bandlimited signals. At the same time,
the condition $\omega<\Lambda_{\ccalS^c}^{\max}$ emphasizes that looking for
a good sampling set is equivalent to pursuing a sampling set whose complement
leads to the maximum possible removable constant.


\begin{remark}
\label{rem_max_vs_min}\normalfont
The appearance of $\Lambda^{\max}$ in
Theorem~\ref{thm_unique_set} deserves emphasis, as it captures one of the
distinctive and somewhat surprising features of sampling on multigraphs relative to sampling on any single
constituent graph: \emph{one informative layer suffices}. A layer that is
poorly connected on $\ccalS^{c}$ -- or even degenerate in the sense of
Remark~\ref{rem_degenerate_sets} -- does not alter the nature of the sampling problem
on the multigraph, as long as some other layer separates $L_{2}(\ccalS^{c})$
from the low frequencies. We also note that the numerical values
of $\Lambda_{r,\ccalS}$ for different $r$ live on different spectral scales
-- those of the respective shift operators -- so that comparisons across
axes, such as the ones implicit
in~\eqref{eq_def_lambda_min_max}, must be interpreted with care. A
scale-free treatment of this issue is given in
Section~\ref{sec_coupled_optimality}.
\end{remark}




%

\section{Coupled Optimality of Sampling Sets in Graphs and Multigraphs}
\label{sec_coupled_optimality}

Given the conditions for uniqueness stated in Theorem~\ref{thm_unique_set},
it is clear that finding a good sampling set $\ccalS$ amounts to finding a
set whose complement leads to the maximum possible value of the removable
constant, and while this interpretation is direct for graphs~\cite{pesensonams1,ortega_proxies,gphon_samp,alejopm_phdthesis}, the
multigraph setting requires a deeper discussion. The removable constants
$\Lambda_{r,\ccalS^{c}}$ are computed \emph{per layer}, and one may
therefore expect the identity of an optimal sampling set to depend on which
layer, or which combination of layers, drives the constants. The central
observation of this section is that this expectation is only partially
correct as the \emph{exact} optimality of a sampling set is a property of the
common Fourier basis $\mathbf{U}$ and the partition $\tau_{N}$ alone -- it
does not depend on any particular constituent graph -- while the layers
re-enter the picture only through the \emph{computable estimates} of that
optimality. We first make the layer-free statement precise, then examine
what survives of it at the level of the constants, and finally exploit the
gap between the two to design a sampling algorithm.


\subsection{Optimality Is a Property of the Common Fourier Basis}
\label{subsec_layer_free}

By Theorem~\ref{thm_unique_set} and the characterization
$\mathcal{PW}_{\omega}(\ccalM)=\operatorname{range}(\mathbf{U}_{\ccalR(\omega)})$
in~\eqref{eq_pw_as_range}, whether a set $\ccalS$ is a
uniqueness set for $\omega$-bandlimited signals is a statement about the
relative position of the two subspaces $L_{2}(\ccalS^{c})$ and
$\operatorname{range}(\mathbf{U}_{\ccalR(\omega)})$. Neither subspace
involves the values of the blocks $\mathbf{\Xi}_{\ell}^{(r)}$. In particular, the former is
determined by $\ccalS$, and the latter by $\mathbf{U}$, the partition
$\tau_{N}$, and the ordering~\eqref{eq_freq_ordering}. This motivates the
following layer-free index of quality for a sampling set.


\begin{definition}[Spectral reach]\label{def_spectral_reach}
Let $\ccalM$ be a multigraph with common Fourier basis $\mathbf{U}$,
partition $\tau_{N}=(q_{1},\ldots,q_{L})$, and blocks ordered as
in~\eqref{eq_freq_ordering}. For $\ccalS\subseteq\ccalV$ with
$\ccalS^{c}\neq\emptyset$, the \emph{spectral reach} of $\ccalS$ is
\begin{equation}\label{eq_def_spectral_reach}
\ell^{*}(\ccalS)
   =
   \min
   \left\lbrace
       \ell
       \; : \;
       L_{2}(\ccalS^{c})
       \cap
       \bigoplus_{j=1}^{\ell}
           \operatorname{range}\left(\mathbf{U}_{j}\right)
       \neq
       \{\mathbf{0}\}
   \right\rbrace
   ,
\end{equation}
with the convention $\ell^{*}(\ccalV)=L+1$ and
$\mu_{L+1}=+\infty$.
\end{definition}


\noindent In words, $\ell^{*}(\ccalS)$ is the index of the first block of~\eqref{eq_pw_as_range} in which a signal supported outside
the samples can \emph{hide}. The minimum in~\eqref{eq_def_spectral_reach} is
well defined because at $\ell=L$ the direct sum equals $\mbR^{N}$, which
intersects $L_{2}(\ccalS^{c})\neq\{\mathbf{0}\}$ nontrivially.
Equivalently, in terms of rank conditions,
\begin{equation}\label{eq_reach_rank}
\ell^{*}(\ccalS)
   =
   \min
   \left\lbrace
       \ell
       \; : \;
       \operatorname{rank}
          \left(
              \mathbf{U}_{\ccalS,\,\tau(1:\ell)}
          \right)
       <
       \textstyle\sum_{j=1}^{\ell} q_{j}
   \right\rbrace
   ,
\end{equation}
where $\tau(1:\ell)$ collects the column indices of the first $\ell$
blocks. The next theorem shows that the spectral reach characterizes
uniqueness sets \emph{exactly}, turning the sufficient condition of
Theorem~\ref{thm_unique_set} into a necessary and sufficient one, at the
price of requiring spectral knowledge.


\begin{theorem}
\label{thm_reach_characterization}
Let $\ccalM$ be a multigraph with blocks ordered as
in~\eqref{eq_freq_ordering}, and let $\ccalS\subseteq\ccalV$. Then $\ccalS$
is a uniqueness set for $\mathcal{PW}_{\omega}(\ccalM)$ if and only if
\begin{equation}\label{eq_thm_reach}
\omega
   <
   \mu_{\ell^{*}(\ccalS)}
   .
\end{equation}
Consequently, for a sampling budget $m$, the sets that are uniqueness sets
for the largest possible family of bandlimited spaces are exactly the
maximizers of $\mu_{\ell^{*}(\cdot)}$, which coincide with the maximizers of
the spectral reach when the labels $\mu_{\ell}$ are pairwise distinct,
\begin{equation}\label{eq_opt_reach}
\ccalS^{\mathrm{opt}}
   \in
   \argmax_{\vert\ccalS\vert = m}
       \ell^{*}(\ccalS)
   =
   \argmax_{\vert\ccalS\vert = m}
       \mu_{\ell^{*}(\ccalS)}
   .
\end{equation}
\end{theorem}

\begin{proof}
    See Appendix~\ref{proof_thm_reach_characterization}.
\end{proof}

\noindent Theorem~\ref{thm_reach_characterization} is the rigorous form of
the following informal statement: \emph{the optimality of a sampling set is
evaluated directly on $\mathbf{U}$}, since at the end of the day any
candidate set is tested against the same basis and the same collection of
invariant subspaces, so its optimality cannot depend on any constituent
graph individually. The role of the individual layers is to provide
\emph{rulers} -- via their removable constants -- with which the layer-free
quantity $\ell^{*}(\ccalS)$ can be estimated without computing
$\mathbf{U}$. To formalize in what sense different rulers certify the same
optimal sets, we introduce the following class of elements of the filter
algebra.


\begin{definition}\label{def_admissible}
Let $\ccalM$ be a multigraph with blocks ordered as
in~\eqref{eq_freq_ordering}. A symmetric filter
$\mathbf{T}=p(\mathbf{S}_{1},\ldots,\mathbf{S}_{R})\in
\ccalA(\mathbf{S}_{1},\ldots,\mathbf{S}_{R})$, with blocks
$\mathbf{\Theta}_{\ell}=p\left( \mathbf{\Xi}_{\ell}^{(1)},\ldots,
\mathbf{\Xi}_{\ell}^{(R)}\right)$, is called \emph{admissible} if its block
spectra are separated consistently with the ordering
of~\eqref{eq_freq_ordering}, i.e., if
\begin{equation}\label{eq_def_admissible}
\max_{\lambda\in\operatorname{spec}(\mathbf{\Theta}_{\ell})}
    \vert\lambda\vert
    \;<\;
\min_{\lambda\in\operatorname{spec}(\mathbf{\Theta}_{\ell+1})}
    \vert\lambda\vert
    ,
    \qquad
    \ell=1,\ldots,L-1
    .
\end{equation}
In particular, a constituent shift operator $\mathbf{S}_{r}$ (when
symmetric) is admissible if its own block spectra are separated and ordered
consistently with~\eqref{eq_freq_ordering}.
\end{definition}


\noindent Since $\mathbf{T}$ shares the basis $\mathbf{U}$ with the shift
operators, its eigenvalues are those of the blocks
$\mathbf{\Theta}_{1},\ldots,\mathbf{\Theta}_{L}$, and
condition~\eqref{eq_def_admissible} requires these groups of eigenvalues to
occupy disjoint intervals of the $\vert\lambda\vert$-axis, arranged in the
same order as the blocks in~\eqref{eq_freq_ordering}. Consequently,
ordering the eigenvalues of $\mathbf{T}$ by magnitude -- as is customary in
GSP -- reproduces the block ordering of $\ccalM$, and every low-pass
subspace of $\mathbf{T}$ whose cutoff falls between two consecutive
intervals is a bandlimited space of the multigraph. Admissibility is a
genuine restriction: the identity matrix, for instance, belongs to the
algebra but is not admissible, since all its blocks share the same
spectrum. An admissible $\mathbf{T}$ can therefore play the role of the
shift operator of a single graph whose bandlimited spaces are those of the
multigraph, and a set that is optimal for sampling on this graph should
also be optimal for $\ccalM$. The following corollary shows that this is
indeed the case: the cutoff frequency computed with respect to any
admissible $\mathbf{T}$ ranks sampling sets exactly as the spectral reach
$\ell^{*}(\cdot)$ of Definition~\ref{def_spectral_reach} does.


\begin{corollary}\label{cor_coupled_opt}
Let $\ccalM$ be a multigraph whose shift operators are jointly block
diagonalized by $\mathbf{U}$ as in~\eqref{eq_jbd}, and let
$\mathbf{T}=\mathbf{U}\operatorname{diag}(\mathbf{\Theta}_{1},\ldots,
\mathbf{\Theta}_{L})\mathbf{U}^{\top}$ be an admissible element of the filter
algebra of $\ccalM$. For each $\ell$, let
$\mathbf{\Theta}_{\ell}=\mathbf{V}_{\ell}\mathbf{D}_{\ell}
\mathbf{V}_{\ell}^{\top}$ be an eigendecomposition of the symmetric block
$\mathbf{\Theta}_{\ell}$, with $\mathbf{V}_{\ell}\in\mbR^{q_{\ell}\times
q_{\ell}}$ orthonormal, and let
\begin{equation}\label{eq_UT}
\mathbf{U}_{\mathbf{T}}
   =
   \mathbf{U}\,
   \operatorname{diag}
      \left(
          \mathbf{V}_{1},\ldots,\mathbf{V}_{L}
      \right)
      ,
\end{equation}
with $\mathbf{T}
   =
   \mathbf{U}_{\mathbf{T}}\,
   \operatorname{diag}
      \left(
          \mathbf{D}_{1},\ldots,\mathbf{D}_{L}
      \right)
   \mathbf{U}_{\mathbf{T}}^{\top}
   .$
We regard $\mathbf{T}$ as the shift operator of a single graph signal model
whose graph Fourier transform is given by $\mathbf{U}_{\mathbf{T}}^{\top}$, and we
denote by $\omega_{\mathbf{T}}(\boldsymbol{\phi})$ the bandwidth of a signal
$\boldsymbol{\phi}$ in this model, i.e., the largest $\vert\lambda\vert$
among the eigenvalues of $\mathbf{T}$ whose corresponding components of
$\mathbf{U}_{\mathbf{T}}^{\top}\boldsymbol{\phi}$ are nonzero. For
$\ccalS\subseteq\ccalV$, let
\begin{equation}\label{eq_cutoff_T}
\nu_{\mathbf{T}}(\ccalS)
   =
   \min_{\boldsymbol{\phi}\in L_{2}(\ccalS^{c}),\
   \boldsymbol{\phi}\neq\mathbf{0}}
       \omega_{\mathbf{T}}(\boldsymbol{\phi})
\end{equation}
be the cutoff frequency of $\ccalS$ with respect to $\mathbf{T}$. Then:
\begin{enumerate}
\item[(i)] $\nu_{\mathbf{T}}(\ccalS)$ lies in the $\ell^{*}(\ccalS)$-th
spectral interval of $\mathbf{T}$, i.e.,
$\nu_{\mathbf{T}}(\ccalS)\in
\left[
\min_{\lambda\in\operatorname{spec}(\mathbf{\Theta}_{\ell^{*}(\ccalS)})}
\vert\lambda\vert,\;
\max_{\lambda\in\operatorname{spec}(\mathbf{\Theta}_{\ell^{*}(\ccalS)})}
\vert\lambda\vert
\right]$;
\item[(ii)] for any two sets $\ccalS_{1},\ccalS_{2}$, if
$\ell^{*}(\ccalS_{1})<\ell^{*}(\ccalS_{2})$ then
$\nu_{\mathbf{T}}(\ccalS_{1})<\nu_{\mathbf{T}}(\ccalS_{2})$;
\item[(iii)] consequently, every maximizer of
$\nu_{\mathbf{T}}(\cdot)$ over $\{\ccalS : \vert\ccalS\vert=m\}$ is a
maximizer of the spectral reach $\ell^{*}(\cdot)$, i.e., an optimal sampling
set of the multigraph in the sense of~\eqref{eq_opt_reach}.
\end{enumerate}
In particular, the optimal sampling sets certified by any two admissible
elements of the algebra coincide at the level of the spectral reach, and
they coincide with the optimal sampling sets of the multigraph.
\end{corollary}

\begin{proof}
    See Appendix~\ref{proof_cor_coupled_opt}.
\end{proof}


\begin{remark}[The Fourier basis of $\mathbf{T}$ versus the multigraph
Fourier basis]\label{rem_UT_vs_U}\normalfont
Corollary~\ref{cor_coupled_opt} involves two orthonormal bases, and it is
worth making their relationship explicit. The matrix $\mathbf{U}$ is the
basis of the joint block diagonalization~\eqref{eq_jbd}: it block
diagonalizes $\mathbf{T}$, but since the blocks $\mathbf{\Theta}_{\ell}$ are
in general not diagonal, its columns are in general \emph{not} eigenvectors
of $\mathbf{T}$. The matrix $\mathbf{U}_{\mathbf{T}}$ in~\eqref{eq_UT} is an
eigenbasis of $\mathbf{T}$, and it is the Fourier basis of $\mathbf{T}$ when
$\mathbf{T}$ is regarded as the shift operator of a graph. The two bases
differ only by the orthogonal rotations $\mathbf{V}_{\ell}$, each acting
\emph{inside} one block. Since these rotations never move energy from one
block to another,
\begin{equation}\label{eq_UT_blocks}
\operatorname{range}
   \left(
       \mathbf{U}_{\mathbf{T}}\,
       \mathbf{I}_{\tau(\ell)}
   \right)
   =
   \operatorname{range}
      \left(
          \mathbf{U}_{\ell}
      \right)
      ,
   \qquad
   \ell=1,\ldots,L
   ,
\end{equation}
where $\mathbf{I}_{\tau(\ell)}$ selects the columns associated with the
$\ell$th block. That is, $\mathbf{U}_{\mathbf{T}}$ and $\mathbf{U}$ define
the same invariant subspaces, and therefore the same $\bigoplus_{j\leq\ell}\operatorname{range}(\mathbf{U}_{j})$ appearing
in~\eqref{eq_pw_as_range} and in Definition~\ref{def_spectral_reach}.

The role of admissibility is to make this agreement visible \emph{from the
spectrum of $\mathbf{T}$ alone}. In the graph signal model of $\mathbf{T}$,
the eigenvectors are ordered by $\vert\lambda\vert$, and the bandlimited
spaces are spanned by the eigenvectors with $\vert\lambda\vert$ below a
threshold. Under~\eqref{eq_def_admissible}, the spectral intervals of the
blocks are disjoint and increasing in $\ell$, so ordering the eigenvalues of
$\mathbf{T}$ by $\vert\lambda\vert$ reproduces exactly the block ordering
of~\eqref{eq_freq_ordering}: the eigenvectors of $\mathbf{T}$ whose
eigenvalues lie in the $\ell$th interval span precisely
$\operatorname{range}(\mathbf{U}_{\ell})$. Consequently, every bandlimited
space of the graph model of $\mathbf{T}$ whose threshold falls in a gap
between two intervals coincides with a bandlimited space
$\mathcal{PW}_{\omega}(\ccalM)$ of the multigraph, and the quantities
$\omega_{\mathbf{T}}$ and $\nu_{\mathbf{T}}$ only interact with
$\mathbf{U}_{\mathbf{T}}$ through this block-level information. This also
shows that $\omega_{\mathbf{T}}(\boldsymbol{\phi})$, and hence
$\nu_{\mathbf{T}}(\ccalS)$, does not depend on the choice of the
$\mathbf{V}_{\ell}$, which is not unique when a block
$\mathbf{\Theta}_{\ell}$ has repeated eigenvalues: by the disjointness of the
intervals, each eigenspace of $\mathbf{T}$ is contained in a single
$\operatorname{range}(\mathbf{U}_{\ell})$, and
$\omega_{\mathbf{T}}(\boldsymbol{\phi})$ is determined by which eigenspaces
$\boldsymbol{\phi}$ has a nonzero projection onto.

If $\mathbf{T}$ is not admissible, the bases $\mathbf{U}_{\mathbf{T}}$ and
$\mathbf{U}$ are still related as in~\eqref{eq_UT}, so the invariant
subspaces still agree, but the spectral intervals of different blocks
overlap. Ordering the eigenvalues of $\mathbf{T}$ by $\vert\lambda\vert$
then \emph{interleaves} eigenvectors coming from different blocks, the
bandlimited spaces of the graph model of $\mathbf{T}$ are no longer
bandlimited spaces of the multigraph, and $\nu_{\mathbf{T}}$ may rank
sampling sets in an order that is unrelated to the spectral reach
$\ell^{*}$. In summary, the coupling expressed by
Corollary~\ref{cor_coupled_opt} is not a coincidence between two different
bases: it is the statement that, under admissibility, the graph Fourier
basis of $\mathbf{T}$ and the multigraph Fourier basis induce the same
ordered filtration of invariant subspaces, and optimality of a sampling set
depends on nothing else.
\end{remark}


This result highlights the connection between the algorithms used to find
optimal sampling sets in graphs and those that can be developed for
multigraphs: \emph{any} admissible single-operator surrogate certifies the
multigraph-optimal sets. Two caveats, however, are essential, and they shape
the algorithmic development of
Section~\ref{subsec_multigraph_algorithm}.


\begin{remark}[Why the coupling cannot be stated at the level of the
constants]\label{rem_maxmin_gap}\normalfont
It is tempting to formulate Corollary~\ref{cor_coupled_opt} directly in
terms of the removable constants of Definition~\ref{def_remov_set}, e.g., by
claiming that a maximizer of
$\Lambda_{\ccalS^{c}}^{\min}=\min_{r}\Lambda_{r,\ccalS^{c}}$ must be a
maximizer of $\Lambda_{r,\ccalS^{c}}$ for some individual layer $r$. Such a
statement is false in general, due to the usual gap between max--min and
min--max: maximizers of a minimum are typically \emph{compromise} solutions
that maximize none of the individual objectives. Schematically, with two
layers and three candidate sets $\ccalA,\ccalB,\ccalC$ of the same size,
constants $(\Lambda_{1,\cdot})=(3,1,2)$ and $(\Lambda_{2,\cdot})=(1,3,2)$
give the max--min optimum at $\ccalC$, while layer $1$ prefers $\ccalA$ and
layer $2$ prefers $\ccalB$. The obstruction is that the constants
$\Lambda_{r,\ccalS}$ measure the common, layer-free quantity
$\ell^{*}(\ccalS)$ on \emph{different spectral scales} -- those of the
respective layers -- and, at first order, they are additionally distorted by
the non-normal, finite-order effects quantified in
Section~\ref{subsec_graph_removable}. The coupling of optimality across
layers is exact at the level of the subspace invariant
$\ell^{*}(\ccalS)$ (Corollary~\ref{cor_coupled_opt}), not at the level of
the numerical constants.
\end{remark}



\begin{remark}[The unknown-graph obstruction]\label{rem_unknown_graph}
\normalfont
Corollary~\ref{cor_coupled_opt} suggests a direct algorithmic route: pick an
admissible element $\mathbf{T}$, and run a single-graph sampling algorithm
on it. The difficulty is that admissibility,
condition~\eqref{eq_def_admissible}, is a spectral property: verifying it
requires precisely the joint block diagonalization whose computation we wish
to avoid, and, generically, \emph{none} of the constituent shift operators
need be admissible -- the layers can order and interleave the block spectra
in incompatible ways. In other words, the optimal sampling set can indeed be
found by working on a specific single operator, but which operators qualify
cannot be known a priori. The resolution, developed in
Section~\ref{subsec_multigraph_algorithm}, is that the \emph{sufficient}
conditions for uniqueness -- unlike the exact optimality statements -- do
not require admissibility at all: they hold unconditionally for a canonical,
computable aggregate of the layers.
\end{remark}



\subsection{Optimal Sampling in Graphs through Removable Constants}
\label{subsec_graph_removable}

Equation~\eqref{eq_lambda_eig}, in the terminology
of~\cite{ortega_proxies}, states that $\Lambda_{r,\ccalS^{c}}$ is the first
order estimate of the cutoff frequency of the sampling set $\ccalS$ on the
graph $(\ccalV, \ccalE_{r},w_r)$: in the notation
of~\cite{ortega_proxies}, $\Lambda_{r,\ccalS^{c}}=\Omega_{1}(\ccalS)$
computed with the variation operator $\mathbf{S}_{r}$. As indicated
in~\cite{ortega_proxies}, this quantity is obtained as the square root of
the smallest eigenvalue of
$\left[\mathbf{S}_{r}^\top \mathbf{S}_{r}\right]_{\ccalS^{c},\ccalS^{c}}$, and the
estimate is refined by taking powers: the $k$th order constants
\begin{equation}\label{eq_lambda_order_k}
\Lambda_{r,\ccalS^{c}}^{(k)}
   =
   \left[
   \lambda_{\min}
      \!\left(
          \left[
              \left(\mathbf{S}_{r}^{k}\right)^{\!\top}
              \mathbf{S}_{r}^{k}
          \right]_{\ccalS^{c},\ccalS^{c}}
       \right)
   \right]^{\frac{1}{2k}}
\end{equation}
converge, as $k\to\infty$, to the smallest layer-$r$ bandwidth attainable by
a signal supported on $\ccalS^{c}$, i.e., to the exact cutoff
frequency~\eqref{eq_cutoff_T} with $\mathbf{T}=\mathbf{S}_{r}$. This
observation is leveraged in~\cite{ortega_proxies} to introduce one of the
iconic sampling schemes in GSP: the optimal sampling set is approximated by
a greedy algorithm that adds one node at a time to $\ccalS$, starting from
$\ccalS=\emptyset$. At each step, the eigenvector associated with the
smallest eigenvalue of
$\left[(\mathbf{S}_{r}^{k})^{\top}\mathbf{S}_{r}^{k}\right]_{\ccalS^{c},\ccalS^{c}}$ --
the \emph{smoothest} signal supported outside the current samples -- is
computed, and its largest component in absolute value determines the node to
be added to $\ccalS$. The greedy step is justified by a binary relaxation of
the support constraint, whose gradient with respect to the relaxed indicator
concentrates exactly on the squared entries of the smoothest
signal~\cite{ortega_proxies}.

From Corollary~\ref{cor_coupled_opt} we can infer that, if one knew an
\emph{admissible} operator for the multigraph -- in particular, if one knew
which constituent graph, if any, is admissible -- then the optimal sampling
set of the multigraph could be determined by directly running the algorithm
of~\cite{ortega_proxies} on that operator. However, as discussed in
Remark~\ref{rem_unknown_graph}, such knowledge is infeasible in a real
problem: it presupposes the spectral information that sampling-without-GFT
methods are designed to avoid. The next subsection shows that this
obstruction can be dissolved rather than circumvented.


\subsection{A Sampling Algorithm for Signals on Multigraphs}
\label{subsec_multigraph_algorithm}

The key observation is that we do not need to identify a distinguished
layer, because we can \emph{construct} an operator that aggregates all the
layers and, by design, lives in the same algebra -- hence shares the basis
$\mathbf{U}$ and the block structure -- while being computable directly from
the shift operators. For $k\in\mathbb{N}$, define the \emph{aggregate
operator of order $k$} and the associated cutoff estimate
\begin{equation}\label{eq_def_Qk}
\mathbf{Q}_{k}
   =
   \sum_{r=1}^{R}
       \left(\mathbf{S}_{r}^{k}\right)^{\!\top}
       \mathbf{S}_{r}^{k}
       ,
\qquad
\Gamma_{k,\ccalS}
   =
   \left[
       \frac{1}{R}\,
       \lambda_{\min}
          \!\left(
              \left[\mathbf{Q}_{k}\right]_{\ccalS,\ccalS}
           \right)
   \right]^{\frac{1}{2k}}
   .
\end{equation}
The operator $\mathbf{Q}_{k}$ is symmetric and positive semidefinite, it is
computable without any spectral information, and it belongs to the algebra
generated by the shift operators and their transposes. We can see this by taking into account that by~\eqref{eq_jbd},
\begin{multline}\label{eq_Qk_blocks}
\mathbf{U}^{\top}\mathbf{Q}_{k}\,\mathbf{U}
   =
   \\
   \operatorname{diag}
      \left(
          \sum_{r=1}^{R}
              \left(\mathbf{\Xi}_{1}^{(r)\,k}\right)^{\!\top}
              \mathbf{\Xi}_{1}^{(r)\,k}
          ,\;
          \ldots
          ,\;
          \sum_{r=1}^{R}
              \left(\mathbf{\Xi}_{L}^{(r)\,k}\right)^{\!\top}
              \mathbf{\Xi}_{L}^{(r)\,k}
      \right)
      ,
\end{multline}
so that $\mathbf{Q}_{k}$ pools, block by block, the spectral information of
all the layers. The following theorem shows that the estimates
$\Gamma_{k,\ccalS^{c}}$ certify uniqueness \emph{unconditionally}: no
admissibility, no knowledge of a distinguished layer, and no monotonicity in
$k$ are required, since each order provides an independent sufficient
condition.


\begin{theorem}
\label{thm_unique_aggregate}
Let $\ccalM = (\ccalV, \{\ccalE_r,w_{r}\}_{r=1}^{R})$ be a multigraph, let
$\ccalS\subset\ccalV$, and let $k\in\mathbb{N}$. Let $(\ccalM,\mathbf{x})$
and $(\ccalM,\mathbf{y})$ be $\omega$-bandlimited signals on $\ccalM$. If
\begin{equation}\label{eq_thm_unique_aggregate}
\omega
   <
   \Gamma_{k,\ccalS^{c}}
\end{equation}
and $\mathbf{x}\vert_{\ccalS}=\mathbf{y}\vert_{\ccalS}$, then
$\mathbf{x}=\mathbf{y}$. Consequently, $\ccalS$ is a uniqueness set for
$\mathcal{PW}_{\omega}(\ccalM)$ whenever
$\omega<\max_{k\in\ccalK}\Gamma_{k,\ccalS^{c}}$ for any finite collection
of orders $\ccalK\subset\mathbb{N}$.
\end{theorem}

\begin{proof}
    See Appendix~\ref{proof_thm_unique_aggregate}.
\end{proof}

The next proposition compares the aggregate estimate with the per-layer
constants of Definition~\ref{def_remov_set}, at matching order $k=1$. It
shows that aggregation is provably \emph{at least as good as the worst
layer} -- strictly better whenever the layers disagree -- while requiring a
single eigenvalue computation instead of $R$ of them. In this precise sense,
the multigraph structure helps the sampling problem: the aggregate is never
held hostage by a degenerate or poorly connected layer.


\begin{proposition}[Domination of the aggregated constants]
\label{prop_domination}
For every nonempty $\ccalS\subset\ccalV$,
\begin{equation}\label{eq_prop_domination}
\frac{\Lambda_{\ccalS}^{\max}}{\sqrt{R}}
\;\leq\;
\Gamma_{1,\ccalS}
\qquad
\text{and}
\qquad
\Lambda_{\ccalS}^{\min}
\;\leq\;
\Gamma_{1,\ccalS}
,
\end{equation}
and the second inequality is strict whenever the layer constants
$\Lambda_{1,\ccalS},\ldots,\Lambda_{R,\ccalS}$ are not all equal.
\end{proposition}

\begin{proof}
    See Appendix~\ref{proof_prop_domination}.
\end{proof}

Since $\mathbf{Q}_{k}$ is symmetric, the classical eigenvalue interlacing
property yields the monotonicity of the cutoff estimates along nested
sampling sets, exactly as in the single-graph
case~\cite{ortega_proxies}. This is the property that makes a greedy
construction meaningful.


\begin{proposition}[Monotonicity]\label{prop_monotonicity}
Let $\ccalS_{1}\subseteq\ccalS_{2}\subset\ccalV$. Then, for every
$k\in\mathbb{N}$,
\begin{equation}
\Gamma_{k,\ccalS_{1}^{c}}
\;\leq\;
\Gamma_{k,\ccalS_{2}^{c}}
.
\end{equation}
\end{proposition}

\begin{proof}
    See Appendix~\ref{proof_prop_monotonicity}.
\end{proof}

With these ingredients, the sampling scheme is the natural multigraph
counterpart of the greedy algorithm of~\cite{ortega_proxies}, with the
matrix $(\mathbf{L}^{\top})^{k}\mathbf{L}^{k}$ built from the Laplacian
$\mathbf{L}$ of the graph replaced by the aggregate $\mathbf{Q}_{k}$. Starting from $\ccalS=\emptyset$, at each step we compute
the eigenvector $\boldsymbol{\psi}_{k}^{*}$ associated with the smallest
eigenvalue of $[\mathbf{Q}_{k}]_{\ccalS^{c},\ccalS^{c}}$ -- the smoothest signal, in
the aggregate sense, supported outside the current samples -- and we add to
$\ccalS$ the node where $\boldsymbol{\psi}_{k}^{*}$ carries maximum energy.
The greedy step is justified verbatim by the binary relaxation argument
of~\cite{ortega_proxies}. The derivation there only requires the reduced
matrix to be symmetric positive semidefinite, which $[\mathbf{Q}_{k}]
_{\ccalS^{c},\ccalS^{c}}$ is, and the gradient of the relaxed objective with respect to
the indicator again concentrates on the squared entries of
$\boldsymbol{\psi}_{k}^{*}$. The procedure is summarized in
Algorithm~\ref{alg_multigraph_sampling}.


\begin{algorithm}[t]
\caption{Greedy sampling set selection on a multigraph}
\label{alg_multigraph_sampling}
\begin{algorithmic}[1]
\Require Shift operators $\{\mathbf{S}_{r}\}_{r=1}^{R}$, sampling budget
$M$, order $k\in\mathbb{N}$.
\Ensure Sampling set $\ccalS$ with $\vert\ccalS\vert=M$.
\State $\ccalS\gets\emptyset$
\While{$\vert\ccalS\vert<M$}
    \State Compute the smallest eigenpair
    $\left(\lambda_{\min},\boldsymbol{\psi}_{k}^{*}\right)$ of
    $\left[\mathbf{Q}_{k}\right]_{\ccalS^{c},\ccalS^{c}}$
    \Comment{matrix-free, see~\eqref{eq_matfree}}
    \State $v\gets\argmax_{i\in\ccalS^{c}}
    \left(\boldsymbol{\psi}_{k}^{*}(i)\right)^{2}$
    \State $\ccalS\gets\ccalS\cup\{v\}$
\EndWhile
\State \Return $\ccalS$
\end{algorithmic}
\end{algorithm}


\noindent\textbf{Complexity.} As in~\cite{ortega_proxies}, the matrix
$[\mathbf{Q}_{k}]_{\ccalS^{c},\ccalS^{c}}$ is never formed explicitly. The step
of any iterative smallest-eigenpair solver is the matrix--vector product
\begin{equation}\label{eq_matfree}
\left[\mathbf{Q}_{k}\right]_{\ccalS^{c},\ccalS^{c}}\mathbf{z}
   =
   \mathbf{I}_{\ccalS^{c}\ccalV}
   \sum_{r=1}^{R}
       \underbrace{
       \mathbf{S}_{r}^{\top}\cdots\mathbf{S}_{r}^{\top}
       }_{k\ \text{times}}
       \underbrace{
       \mathbf{S}_{r}\cdots\mathbf{S}_{r}
       }_{k\ \text{times}}
   \mathbf{I}_{\ccalV\ccalS^{c}}\,
   \mathbf{z}
   ,
\end{equation}
which amounts to $2k$ sparse matrix--vector products per layer, for a cost
of $O\!\left(k\sum_{r=1}^{R}\vert\ccalE_{r}\vert\right)$ per product and
$O(N)$ storage. The per-iteration cost is thus the cost of running the
single-graph method of~\cite{ortega_proxies} on each layer once --
divided by the fact that only \emph{one} eigenpair, of the aggregate, is
computed, instead of one eigenpair per layer as a per-layer implementation
of Definition~\ref{def_remov_set} would require.


\begin{remark}\label{rem_words}\normalfont
The aggregate~\eqref{eq_def_Qk} uses only the powers
$\mathbf{S}_{r}^{k}$, i.e., the constant words. The construction and the
proof of Theorem~\ref{thm_unique_aggregate} extend verbatim to aggregates
over any finite collection $\ccalW$ of words of a common length $k$,
\begin{equation}
\mathbf{Q}_{\ccalW}
   =
   \sum_{w\in\ccalW}
       \mathbf{S}_{w}^{\top}\mathbf{S}_{w}
       ,
\quad
\Gamma_{\ccalW,\ccalS}
   =
   \left[
       \frac{1}{\vert\ccalW\vert}\,
       \lambda_{\min}
          \!\left(
              \left[\mathbf{Q}_{\ccalW}\right]_{\ccalS,\ccalS}
           \right)
   \right]^{\frac{1}{2k}}
   ,
\end{equation}
by Corollary~\ref{cor_word_bernstein}. Mixed words such as
$\mathbf{S}_{2}\mathbf{S}_{1}$ probe the interaction between the layers --
information invisible to any constituent graph in isolation -- and are the
natural candidates for tightening the estimates when the layers are strongly
noncommuting. A systematic study of the choice of $\ccalW$ is left for
future work.
\end{remark}



\begin{remark}\label{rem_blue_note}\normalfont
We can now close the loop with
Section~\ref{subsec_layer_free}. The aggregate $\mathbf{Q}_{k}$ \emph{is}
``the specific graph'' anticipated by Corollary~\ref{cor_coupled_opt}: the
reduction of the multigraph sampling problem to a single-operator sampling
problem is real, but the single operator need not be one of the $R$
constituents, whose admissibility cannot be verified. It can be taken to be
a canonical, computable element of the algebra which shares the basis
$\mathbf{U}$ by construction. At the exact level, optimality is layer-free, i.e.,
for $k\to\infty$ the greedy procedure
of~\cite{ortega_proxies} collapses onto a rank-revealing Gaussian
elimination with partial pivoting on the common eigenvector matrix
(cf.~Proposition~4 of~\cite{ortega_proxies}), so that, when run on any
admissible element of the algebra, it executes the \emph{same} elimination
on the \emph{same} $\mathbf{U}$ -- respecting the block structure -- and
certifies the same optimal sets, in agreement with
Theorem~\ref{thm_reach_characterization}. At finite $k$, the layers
genuinely differ: the quality of a given operator as a ruler for
$\ell^{*}(\ccalS)$ is governed by the separation of its block spectra around
the reach, and~\eqref{eq_Qk_blocks} shows that the aggregate pools the
separation of all the layers, with
Proposition~\ref{prop_domination} quantifying that it never does worse than
the worst-case per-layer construction. The unknown-graph obstruction of
Remark~\ref{rem_unknown_graph} is thus not circumvented but dissolved. More specifically, in exact terms, the choice of operator is irrelevant, and in approximate terms, it is replaced by a computable surrogate.
\end{remark}




%
%
\section{Numerical Experiments and Applications}
\label{sec_experiments}

To evaluate our findings, we conduct a collection of experiments that
consider several multigraphs, classes of bandlimited signals, and sampling
rates. In the experiment, a bandlimited signal on a
multigraph is sampled on a set selected by
Algorithm~\ref{alg_multigraph_sampling}, it is then reconstructed from its
noisy samples, and the reconstruction error is compared with the errors
obtained when the sampling set is instead selected by single-graph methods
applied to the individual layers, by a per-layer worst-case greedy rule, and
at random. In what follows, we describe how the multigraphs and the signals
are generated, the sampling methods under comparison, the methods used
to evaluate them, and we then discuss the results.

\begin{figure*}[t]
	\centering
\providecommand{\aspfigdir}{./figures/experiments}
\ifcsname pgfplotsgroupplotsmath@decrease\endcsname\else\PackageError{asp figures}{pgfplots libraries not loaded}{Add to the PREAMBLE of the main file: \string\usepackage{pgfplots} \string\usepgfplotslibrary{groupplots,fillbetween} \string\usetikzlibrary{calc} \string\pgfplotsset{compat=1.18}}\fi
\ifdefined\aspstyleloaded\else\input{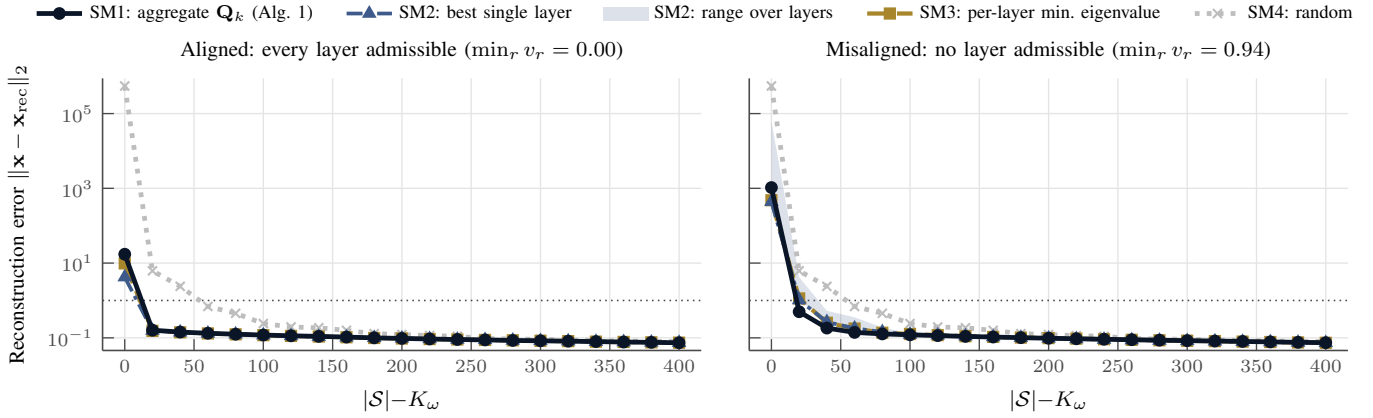}\fi
\begin{tikzpicture}
\pgfmathsetlengthmacro{\aspPW}{0.425*\textwidth}
\pgfmathsetlengthmacro{\aspPH}{\aspPW/2.2}
\begin{groupplot}[
  group style={group name=grp, group size=2 by 1, horizontal sep=0.25in, vertical sep=0.3in, xlabels at=edge bottom, xticklabels at=edge bottom, ylabels at=edge left, yticklabels at=edge left},
  asp axis, asp log, asp legend, width=\aspPW, height=\aspPH,
  xlabel={$\vert\mathcal{S}\vert - K_\omega$}, ylabel={Reconstruction error $\Vert\mathbf{x}-\mathbf{x}_{\mathrm{rec}}\Vert_2$},
]
  \nextgroupplot[title={Aligned: every layer admissible ($\mathrm{min}_r\, v_r = 0.00$)}, xmin=-16, xmax=416, ymin=0.0459447, ymax=876252, xtick={0,50,100,150,200,250,300,350,400}, ytickten={-1,1,3,5}, legend to name=legfig1]
  \addlegendimage{sm1}\addlegendentry{SM1: aggregate $\mathbf{Q}_k$ (Alg.~1)}
  \addlegendimage{sm2best}\addlegendentry{SM2: best single layer}
  \addlegendimage{draw=none, legend image code/.code={\path[draw=none, fill=asplayer, fill opacity=0.18] (0cm,-0.09cm) rectangle (0.6cm,0.09cm);}}\addlegendentry{SM2: range over layers}
  \addlegendimage{sm3}\addlegendentry{SM3: per-layer min.\ eigenvalue}
  \addlegendimage{sm4}\addlegendentry{SM4: random}
  \addplot[name path=lo0, draw=none, forget plot] table[x=x, y=SM2min] {\aspfigdir/data/fig1_aligned.dat};
  \addplot[name path=hi0, draw=none, forget plot] table[x=x, y=SM2max] {\aspfigdir/data/fig1_aligned.dat};
  \addplot[sm2band, forget plot] fill between[of=hi0 and lo0];
  \draw[refline] ({rel axis cs:0,0} |- {axis cs:0,1}) -- ({rel axis cs:1,0} |- {axis cs:0,1});
  \addplot[sm4, forget plot] table[x=x, y=SM4] {\aspfigdir/data/fig1_aligned.dat};
  \addplot[sm3, forget plot] table[x=x, y=SM3] {\aspfigdir/data/fig1_aligned.dat};
  \addplot[sm2best, forget plot] table[x=x, y=SM2best] {\aspfigdir/data/fig1_aligned.dat};
  \addplot[sm1, forget plot] table[x=x, y=SM1] {\aspfigdir/data/fig1_aligned.dat};
  \nextgroupplot[title={Misaligned: no layer admissible ($\mathrm{min}_r\, v_r = 0.94$)}, xmin=-16, xmax=416, ymin=0.0459447, ymax=876252, xtick={0,50,100,150,200,250,300,350,400}, ytickten={-1,1,3,5}]
  \addplot[name path=lo1, draw=none, forget plot] table[x=x, y=SM2min] {\aspfigdir/data/fig1_misaligned.dat};
  \addplot[name path=hi1, draw=none, forget plot] table[x=x, y=SM2max] {\aspfigdir/data/fig1_misaligned.dat};
  \addplot[sm2band, forget plot] fill between[of=hi1 and lo1];
  \draw[refline] ({rel axis cs:0,0} |- {axis cs:0,1}) -- ({rel axis cs:1,0} |- {axis cs:0,1});
  \addplot[sm4, forget plot] table[x=x, y=SM4] {\aspfigdir/data/fig1_misaligned.dat};
  \addplot[sm3, forget plot] table[x=x, y=SM3] {\aspfigdir/data/fig1_misaligned.dat};
  \addplot[sm2best, forget plot] table[x=x, y=SM2best] {\aspfigdir/data/fig1_misaligned.dat};
  \addplot[sm1, forget plot] table[x=x, y=SM1] {\aspfigdir/data/fig1_misaligned.dat};
\end{groupplot}
\node[anchor=south, inner sep=0pt] at ($(grp c1r1.outer north)!0.5!(grp c2r1.outer north)+(0,0pt)$)
  {\pgfplotslegendfromname{legfig1}};
\end{tikzpicture}
	\caption{Reconstruction error as a function of the number of samples
	beyond the bandwidth, $\vert\ccalS\vert-K_{\omega}$, for an aligned
	(left) and a misaligned (right) multigraph with the sensor-graph Fourier
	basis, $R=2$ layers, $K_{\omega}\approx 0.08N$, and
	$\mathrm{SNR}=15\,\mathrm{dB}$. The value of $\min_{r}v_{r}$ in each
	title is the violation fraction~\eqref{eq_exp_violation} of the most
	admissible layer. The shaded band shows the range of SM2 over the
	layers. In all the figures, the error of each realization is averaged
	over $50$ signals, and the median over $5$ realizations is shown. The
	dotted line marks $E=1$.}
	\label{fig_exp_regimes}
\end{figure*}

\begin{figure*}[t]
	\centering
\providecommand{\aspfigdir}{./figures/experiments}
\ifcsname pgfplotsgroupplotsmath@decrease\endcsname\else\PackageError{asp figures}{pgfplots libraries not loaded}{Add to the PREAMBLE of the main file: \string\usepackage{pgfplots} \string\usepgfplotslibrary{groupplots,fillbetween} \string\usetikzlibrary{calc} \string\pgfplotsset{compat=1.18}}\fi
\ifdefined\aspstyleloaded\else\input{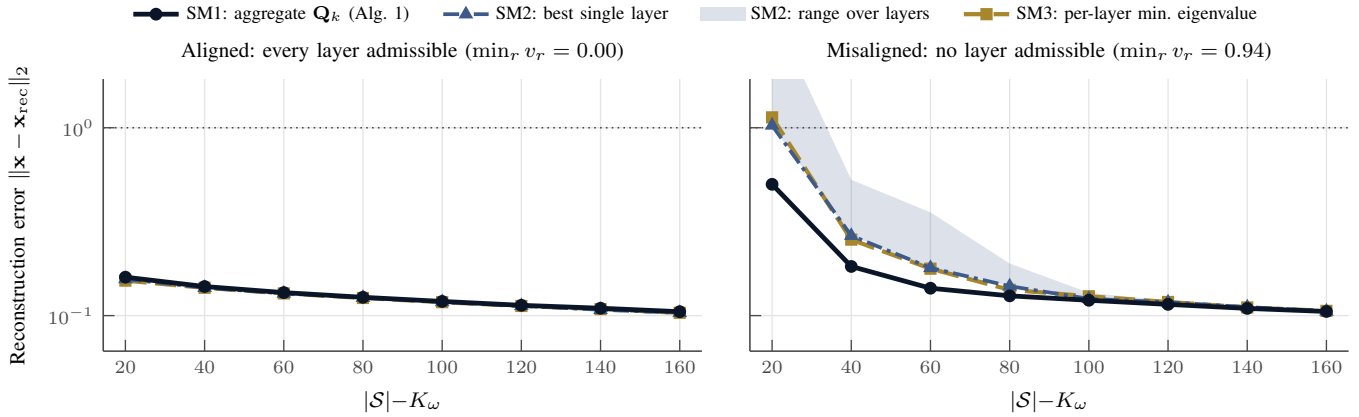}\fi
\begin{tikzpicture}
\pgfmathsetlengthmacro{\aspPW}{0.425*\textwidth}
\pgfmathsetlengthmacro{\aspPH}{\aspPW/2.2}
\begin{groupplot}[
  group style={group name=grp, group size=2 by 1, horizontal sep=0.25in, vertical sep=0.3in, xlabels at=edge bottom, xticklabels at=edge bottom, ylabels at=edge left, yticklabels at=edge left},
  asp axis, asp log, asp legend, width=\aspPW, height=\aspPH,
  xlabel={$\vert\mathcal{S}\vert - K_\omega$}, ylabel={Reconstruction error $\Vert\mathbf{x}-\mathbf{x}_{\mathrm{rec}}\Vert_2$},
]
  \nextgroupplot[title={Aligned: every layer admissible ($\mathrm{min}_r\, v_r = 0.00$)}, xmin=14.4, xmax=165.6, ymin=0.0646879, ymax=1.81709, xtick={20,40,60,80,100,120,140,160}, ytickten={-1,0}, legend to name=legfig2]
  \addlegendimage{sm1}\addlegendentry{SM1: aggregate $\mathbf{Q}_k$ (Alg.~1)}
  \addlegendimage{sm2best}\addlegendentry{SM2: best single layer}
  \addlegendimage{draw=none, legend image code/.code={\path[draw=none, fill=asplayer, fill opacity=0.18] (0cm,-0.09cm) rectangle (0.6cm,0.09cm);}}\addlegendentry{SM2: range over layers}
  \addlegendimage{sm3}\addlegendentry{SM3: per-layer min.\ eigenvalue}
  \addplot[name path=lo0, draw=none, forget plot] table[x=x, y=SM2min] {\aspfigdir/data/fig2_aligned.dat};
  \addplot[name path=hi0, draw=none, forget plot] table[x=x, y=SM2max] {\aspfigdir/data/fig2_aligned.dat};
  \addplot[sm2band, forget plot] fill between[of=hi0 and lo0];
  \draw[refline] ({rel axis cs:0,0} |- {axis cs:0,1}) -- ({rel axis cs:1,0} |- {axis cs:0,1});
  \addplot[sm3, forget plot] table[x=x, y=SM3] {\aspfigdir/data/fig2_aligned.dat};
  \addplot[sm2best, forget plot] table[x=x, y=SM2best] {\aspfigdir/data/fig2_aligned.dat};
  \addplot[sm1, forget plot] table[x=x, y=SM1] {\aspfigdir/data/fig2_aligned.dat};
  \nextgroupplot[title={Misaligned: no layer admissible ($\mathrm{min}_r\, v_r = 0.94$)}, xmin=14.4, xmax=165.6, ymin=0.0646879, ymax=1.81709, xtick={20,40,60,80,100,120,140,160}, ytickten={-1,0}]
  \addplot[name path=lo1, draw=none, forget plot] table[x=x, y=SM2min] {\aspfigdir/data/fig2_misaligned.dat};
  \addplot[name path=hi1, draw=none, forget plot] table[x=x, y=SM2max] {\aspfigdir/data/fig2_misaligned.dat};
  \addplot[sm2band, forget plot] fill between[of=hi1 and lo1];
  \draw[refline] ({rel axis cs:0,0} |- {axis cs:0,1}) -- ({rel axis cs:1,0} |- {axis cs:0,1});
  \addplot[sm3, forget plot] table[x=x, y=SM3] {\aspfigdir/data/fig2_misaligned.dat};
  \addplot[sm2best, forget plot] table[x=x, y=SM2best] {\aspfigdir/data/fig2_misaligned.dat};
  \addplot[sm1, forget plot] table[x=x, y=SM1] {\aspfigdir/data/fig2_misaligned.dat};
\end{groupplot}
\node[anchor=south, inner sep=0pt] at ($(grp c1r1.outer north)!0.5!(grp c2r1.outer north)+(0,0pt)$)
  {\pgfplotslegendfromname{legfig2}};
\end{tikzpicture}
	\caption{Detail of Fig.~\ref{fig_exp_regimes} for
	$20\leq\vert\ccalS\vert-K_{\omega}\leq 160$, without random sampling.
	The critical size $\vert\ccalS\vert=K_{\omega}$, where every method fails,
	is omitted so that the vertical axis resolves the differences between
	the methods; the axis is fitted to the curves of the methods, so the
	range of SM2 over the layers may be clipped.}
	\label{fig_exp_regimes_zoom}
\end{figure*}

\begin{figure*}[t]
	\centering
\providecommand{\aspfigdir}{./figures/experiments}
\ifcsname pgfplotsgroupplotsmath@decrease\endcsname\else\PackageError{asp figures}{pgfplots libraries not loaded}{Add to the PREAMBLE of the main file: \string\usepackage{pgfplots} \string\usepgfplotslibrary{groupplots,fillbetween} \string\usetikzlibrary{calc} \string\pgfplotsset{compat=1.18}}\fi
\ifdefined\aspstyleloaded\else\input{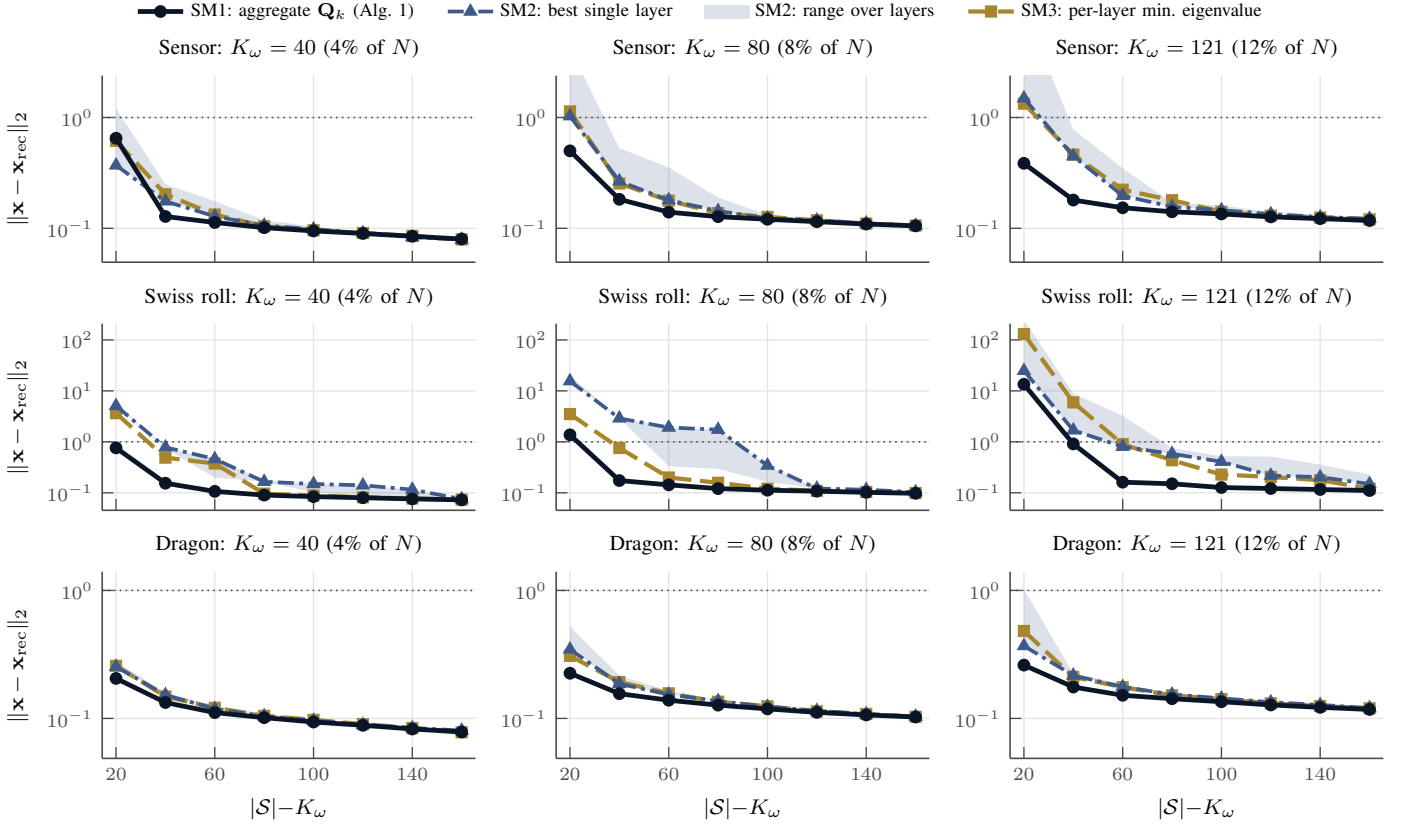}\fi
\begin{tikzpicture}
\pgfmathsetlengthmacro{\aspPW}{0.265*\textwidth}
\pgfmathsetlengthmacro{\aspPH}{\aspPW/2}
\begin{groupplot}[
  group style={group name=grp, group size=3 by 3, horizontal sep=0.42in, vertical sep=0.32in, xlabels at=edge bottom, xticklabels at=edge bottom, ylabels at=edge left},
  asp axis, asp log, asp legend, width=\aspPW, height=\aspPH,
  xlabel={$\vert\mathcal{S}\vert - K_\omega$}, ylabel={$\Vert\mathbf{x}-\mathbf{x}_{\mathrm{rec}}\Vert_2$},
]
  \nextgroupplot[title={Sensor: $K_\omega = 40$ (4\% of $N$)}, xmin=14.4, xmax=165.6, ymin=0.0494661, ymax=2.39023, xtick={20,60,100,140}, ytickten={-1,0}, legend to name=legfig3]
  \addlegendimage{sm1}\addlegendentry{SM1: aggregate $\mathbf{Q}_k$ (Alg.~1)}
  \addlegendimage{sm2best}\addlegendentry{SM2: best single layer}
  \addlegendimage{draw=none, legend image code/.code={\path[draw=none, fill=asplayer, fill opacity=0.18] (0cm,-0.09cm) rectangle (0.6cm,0.09cm);}}\addlegendentry{SM2: range over layers}
  \addlegendimage{sm3}\addlegendentry{SM3: per-layer min.\ eigenvalue}
  \addplot[name path=lo0, draw=none, forget plot] table[x=x, y=SM2min] {\aspfigdir/data/fig3_sensor_bw4.dat};
  \addplot[name path=hi0, draw=none, forget plot] table[x=x, y=SM2max] {\aspfigdir/data/fig3_sensor_bw4.dat};
  \addplot[sm2band, forget plot] fill between[of=hi0 and lo0];
  \draw[refline] ({rel axis cs:0,0} |- {axis cs:0,1}) -- ({rel axis cs:1,0} |- {axis cs:0,1});
  \addplot[sm3, forget plot] table[x=x, y=SM3] {\aspfigdir/data/fig3_sensor_bw4.dat};
  \addplot[sm2best, forget plot] table[x=x, y=SM2best] {\aspfigdir/data/fig3_sensor_bw4.dat};
  \addplot[sm1, forget plot] table[x=x, y=SM1] {\aspfigdir/data/fig3_sensor_bw4.dat};
  \nextgroupplot[title={Sensor: $K_\omega = 80$ (8\% of $N$)}, xmin=14.4, xmax=165.6, ymin=0.0494661, ymax=2.39023, xtick={20,60,100,140}, ytickten={-1,0}]
  \addplot[name path=lo1, draw=none, forget plot] table[x=x, y=SM2min] {\aspfigdir/data/fig3_sensor_bw8.dat};
  \addplot[name path=hi1, draw=none, forget plot] table[x=x, y=SM2max] {\aspfigdir/data/fig3_sensor_bw8.dat};
  \addplot[sm2band, forget plot] fill between[of=hi1 and lo1];
  \draw[refline] ({rel axis cs:0,0} |- {axis cs:0,1}) -- ({rel axis cs:1,0} |- {axis cs:0,1});
  \addplot[sm3, forget plot] table[x=x, y=SM3] {\aspfigdir/data/fig3_sensor_bw8.dat};
  \addplot[sm2best, forget plot] table[x=x, y=SM2best] {\aspfigdir/data/fig3_sensor_bw8.dat};
  \addplot[sm1, forget plot] table[x=x, y=SM1] {\aspfigdir/data/fig3_sensor_bw8.dat};
  \nextgroupplot[title={Sensor: $K_\omega = 121$ (12\% of $N$)}, xmin=14.4, xmax=165.6, ymin=0.0494661, ymax=2.39023, xtick={20,60,100,140}, ytickten={-1,0}]
  \addplot[name path=lo2, draw=none, forget plot] table[x=x, y=SM2min] {\aspfigdir/data/fig3_sensor_bw12.dat};
  \addplot[name path=hi2, draw=none, forget plot] table[x=x, y=SM2max] {\aspfigdir/data/fig3_sensor_bw12.dat};
  \addplot[sm2band, forget plot] fill between[of=hi2 and lo2];
  \draw[refline] ({rel axis cs:0,0} |- {axis cs:0,1}) -- ({rel axis cs:1,0} |- {axis cs:0,1});
  \addplot[sm3, forget plot] table[x=x, y=SM3] {\aspfigdir/data/fig3_sensor_bw12.dat};
  \addplot[sm2best, forget plot] table[x=x, y=SM2best] {\aspfigdir/data/fig3_sensor_bw12.dat};
  \addplot[sm1, forget plot] table[x=x, y=SM1] {\aspfigdir/data/fig3_sensor_bw12.dat};
  \nextgroupplot[title={Swiss roll: $K_\omega = 40$ (4\% of $N$)}, xmin=14.4, xmax=165.6, ymin=0.0453895, ymax=208.447, xtick={20,60,100,140}, ytickten={-1,0,1,2}]
  \addplot[name path=lo3, draw=none, forget plot] table[x=x, y=SM2min] {\aspfigdir/data/fig3_swiss_roll_bw4.dat};
  \addplot[name path=hi3, draw=none, forget plot] table[x=x, y=SM2max] {\aspfigdir/data/fig3_swiss_roll_bw4.dat};
  \addplot[sm2band, forget plot] fill between[of=hi3 and lo3];
  \draw[refline] ({rel axis cs:0,0} |- {axis cs:0,1}) -- ({rel axis cs:1,0} |- {axis cs:0,1});
  \addplot[sm3, forget plot] table[x=x, y=SM3] {\aspfigdir/data/fig3_swiss_roll_bw4.dat};
  \addplot[sm2best, forget plot] table[x=x, y=SM2best] {\aspfigdir/data/fig3_swiss_roll_bw4.dat};
  \addplot[sm1, forget plot] table[x=x, y=SM1] {\aspfigdir/data/fig3_swiss_roll_bw4.dat};
  \nextgroupplot[title={Swiss roll: $K_\omega = 80$ (8\% of $N$)}, xmin=14.4, xmax=165.6, ymin=0.0453895, ymax=208.447, xtick={20,60,100,140}, ytickten={-1,0,1,2}]
  \addplot[name path=lo4, draw=none, forget plot] table[x=x, y=SM2min] {\aspfigdir/data/fig3_swiss_roll_bw8.dat};
  \addplot[name path=hi4, draw=none, forget plot] table[x=x, y=SM2max] {\aspfigdir/data/fig3_swiss_roll_bw8.dat};
  \addplot[sm2band, forget plot] fill between[of=hi4 and lo4];
  \draw[refline] ({rel axis cs:0,0} |- {axis cs:0,1}) -- ({rel axis cs:1,0} |- {axis cs:0,1});
  \addplot[sm3, forget plot] table[x=x, y=SM3] {\aspfigdir/data/fig3_swiss_roll_bw8.dat};
  \addplot[sm2best, forget plot] table[x=x, y=SM2best] {\aspfigdir/data/fig3_swiss_roll_bw8.dat};
  \addplot[sm1, forget plot] table[x=x, y=SM1] {\aspfigdir/data/fig3_swiss_roll_bw8.dat};
  \nextgroupplot[title={Swiss roll: $K_\omega = 121$ (12\% of $N$)}, xmin=14.4, xmax=165.6, ymin=0.0453895, ymax=208.447, xtick={20,60,100,140}, ytickten={-1,0,1,2}]
  \addplot[name path=lo5, draw=none, forget plot] table[x=x, y=SM2min] {\aspfigdir/data/fig3_swiss_roll_bw12.dat};
  \addplot[name path=hi5, draw=none, forget plot] table[x=x, y=SM2max] {\aspfigdir/data/fig3_swiss_roll_bw12.dat};
  \addplot[sm2band, forget plot] fill between[of=hi5 and lo5];
  \draw[refline] ({rel axis cs:0,0} |- {axis cs:0,1}) -- ({rel axis cs:1,0} |- {axis cs:0,1});
  \addplot[sm3, forget plot] table[x=x, y=SM3] {\aspfigdir/data/fig3_swiss_roll_bw12.dat};
  \addplot[sm2best, forget plot] table[x=x, y=SM2best] {\aspfigdir/data/fig3_swiss_roll_bw12.dat};
  \addplot[sm1, forget plot] table[x=x, y=SM1] {\aspfigdir/data/fig3_swiss_roll_bw12.dat};
  \nextgroupplot[title={Dragon: $K_\omega = 40$ (4\% of $N$)}, xmin=14.4, xmax=165.6, ymin=0.0487319, ymax=1.4, xtick={20,60,100,140}, ytickten={-1,0}]
  \addplot[name path=lo6, draw=none, forget plot] table[x=x, y=SM2min] {\aspfigdir/data/fig3_dragon_bw4.dat};
  \addplot[name path=hi6, draw=none, forget plot] table[x=x, y=SM2max] {\aspfigdir/data/fig3_dragon_bw4.dat};
  \addplot[sm2band, forget plot] fill between[of=hi6 and lo6];
  \draw[refline] ({rel axis cs:0,0} |- {axis cs:0,1}) -- ({rel axis cs:1,0} |- {axis cs:0,1});
  \addplot[sm3, forget plot] table[x=x, y=SM3] {\aspfigdir/data/fig3_dragon_bw4.dat};
  \addplot[sm2best, forget plot] table[x=x, y=SM2best] {\aspfigdir/data/fig3_dragon_bw4.dat};
  \addplot[sm1, forget plot] table[x=x, y=SM1] {\aspfigdir/data/fig3_dragon_bw4.dat};
  \nextgroupplot[title={Dragon: $K_\omega = 80$ (8\% of $N$)}, xmin=14.4, xmax=165.6, ymin=0.0487319, ymax=1.4, xtick={20,60,100,140}, ytickten={-1,0}]
  \addplot[name path=lo7, draw=none, forget plot] table[x=x, y=SM2min] {\aspfigdir/data/fig3_dragon_bw8.dat};
  \addplot[name path=hi7, draw=none, forget plot] table[x=x, y=SM2max] {\aspfigdir/data/fig3_dragon_bw8.dat};
  \addplot[sm2band, forget plot] fill between[of=hi7 and lo7];
  \draw[refline] ({rel axis cs:0,0} |- {axis cs:0,1}) -- ({rel axis cs:1,0} |- {axis cs:0,1});
  \addplot[sm3, forget plot] table[x=x, y=SM3] {\aspfigdir/data/fig3_dragon_bw8.dat};
  \addplot[sm2best, forget plot] table[x=x, y=SM2best] {\aspfigdir/data/fig3_dragon_bw8.dat};
  \addplot[sm1, forget plot] table[x=x, y=SM1] {\aspfigdir/data/fig3_dragon_bw8.dat};
  \nextgroupplot[title={Dragon: $K_\omega = 121$ (12\% of $N$)}, xmin=14.4, xmax=165.6, ymin=0.0487319, ymax=1.4, xtick={20,60,100,140}, ytickten={-1,0}]
  \addplot[name path=lo8, draw=none, forget plot] table[x=x, y=SM2min] {\aspfigdir/data/fig3_dragon_bw12.dat};
  \addplot[name path=hi8, draw=none, forget plot] table[x=x, y=SM2max] {\aspfigdir/data/fig3_dragon_bw12.dat};
  \addplot[sm2band, forget plot] fill between[of=hi8 and lo8];
  \draw[refline] ({rel axis cs:0,0} |- {axis cs:0,1}) -- ({rel axis cs:1,0} |- {axis cs:0,1});
  \addplot[sm3, forget plot] table[x=x, y=SM3] {\aspfigdir/data/fig3_dragon_bw12.dat};
  \addplot[sm2best, forget plot] table[x=x, y=SM2best] {\aspfigdir/data/fig3_dragon_bw12.dat};
  \addplot[sm1, forget plot] table[x=x, y=SM1] {\aspfigdir/data/fig3_dragon_bw12.dat};
\end{groupplot}
\node[anchor=south, inner sep=0pt] at ($(grp c1r1.outer north)!0.5!(grp c3r1.outer north)+(0,0pt)$)
  {\pgfplotslegendfromname{legfig3}};
\end{tikzpicture}
	\caption{Reconstruction error on misaligned multigraphs with $R=2$
	layers for the three Fourier bases (rows) and the three bandwidths
	(columns), for $20\leq\vert\ccalS\vert-K_{\omega}\leq 160$ and
	$\mathrm{SNR}=15\,\mathrm{dB}$. Random sampling is omitted; its errors
	are reported in Table~\ref{tab_exp_errors}.}
	\label{fig_exp_bases}
\end{figure*}

\begin{figure*}[t]
	\centering
\providecommand{\aspfigdir}{./figures/experiments}
\ifcsname pgfplotsgroupplotsmath@decrease\endcsname\else\PackageError{asp figures}{pgfplots libraries not loaded}{Add to the PREAMBLE of the main file: \string\usepackage{pgfplots} \string\usepgfplotslibrary{groupplots,fillbetween} \string\usetikzlibrary{calc} \string\pgfplotsset{compat=1.18}}\fi
\ifdefined\aspstyleloaded\else\input{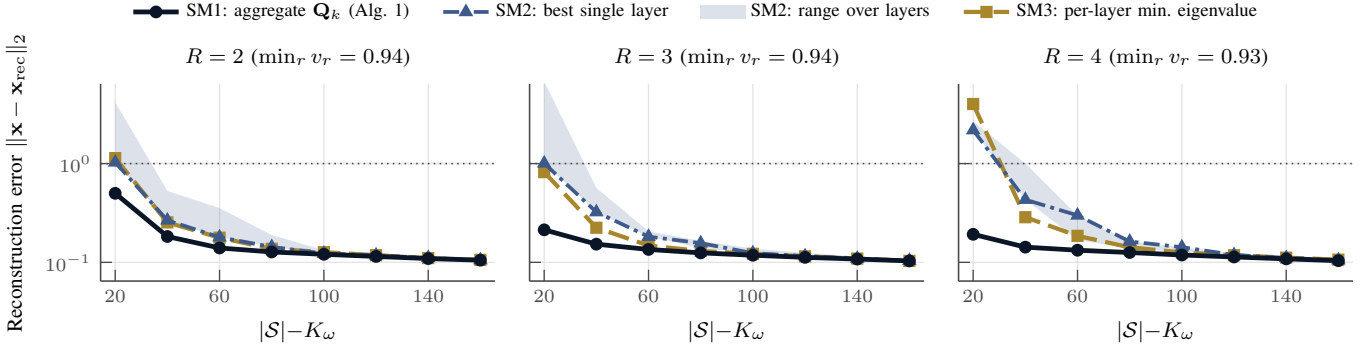}\fi
\begin{tikzpicture}
\pgfmathsetlengthmacro{\aspPW}{0.28*\textwidth}
\pgfmathsetlengthmacro{\aspPH}{\aspPW/2}
\begin{groupplot}[
  group style={group name=grp, group size=3 by 1, horizontal sep=0.18in, vertical sep=0.3in, xlabels at=edge bottom, xticklabels at=edge bottom, ylabels at=edge left, yticklabels at=edge left},
  asp axis, asp log, asp legend, width=\aspPW, height=\aspPH,
  xlabel={$\vert\mathcal{S}\vert - K_\omega$}, ylabel={Reconstruction error $\Vert\mathbf{x}-\mathbf{x}_{\mathrm{rec}}\Vert_2$},
]
  \nextgroupplot[title={$R = 2$ ($\mathrm{min}_r\, v_r = 0.94$)}, xmin=14.4, xmax=165.6, ymin=0.0646427, ymax=6.41475, xtick={20,60,100,140}, ytickten={-1,0}, legend to name=legfig4]
  \addlegendimage{sm1}\addlegendentry{SM1: aggregate $\mathbf{Q}_k$ (Alg.~1)}
  \addlegendimage{sm2best}\addlegendentry{SM2: best single layer}
  \addlegendimage{draw=none, legend image code/.code={\path[draw=none, fill=asplayer, fill opacity=0.18] (0cm,-0.09cm) rectangle (0.6cm,0.09cm);}}\addlegendentry{SM2: range over layers}
  \addlegendimage{sm3}\addlegendentry{SM3: per-layer min.\ eigenvalue}
  \addplot[name path=lo0, draw=none, forget plot] table[x=x, y=SM2min] {\aspfigdir/data/fig4_R2.dat};
  \addplot[name path=hi0, draw=none, forget plot] table[x=x, y=SM2max] {\aspfigdir/data/fig4_R2.dat};
  \addplot[sm2band, forget plot] fill between[of=hi0 and lo0];
  \draw[refline] ({rel axis cs:0,0} |- {axis cs:0,1}) -- ({rel axis cs:1,0} |- {axis cs:0,1});
  \addplot[sm3, forget plot] table[x=x, y=SM3] {\aspfigdir/data/fig4_R2.dat};
  \addplot[sm2best, forget plot] table[x=x, y=SM2best] {\aspfigdir/data/fig4_R2.dat};
  \addplot[sm1, forget plot] table[x=x, y=SM1] {\aspfigdir/data/fig4_R2.dat};
  \nextgroupplot[title={$R = 3$ ($\mathrm{min}_r\, v_r = 0.94$)}, xmin=14.4, xmax=165.6, ymin=0.0646427, ymax=6.41475, xtick={20,60,100,140}, ytickten={-1,0}]
  \addplot[name path=lo1, draw=none, forget plot] table[x=x, y=SM2min] {\aspfigdir/data/fig4_R3.dat};
  \addplot[name path=hi1, draw=none, forget plot] table[x=x, y=SM2max] {\aspfigdir/data/fig4_R3.dat};
  \addplot[sm2band, forget plot] fill between[of=hi1 and lo1];
  \draw[refline] ({rel axis cs:0,0} |- {axis cs:0,1}) -- ({rel axis cs:1,0} |- {axis cs:0,1});
  \addplot[sm3, forget plot] table[x=x, y=SM3] {\aspfigdir/data/fig4_R3.dat};
  \addplot[sm2best, forget plot] table[x=x, y=SM2best] {\aspfigdir/data/fig4_R3.dat};
  \addplot[sm1, forget plot] table[x=x, y=SM1] {\aspfigdir/data/fig4_R3.dat};
  \nextgroupplot[title={$R = 4$ ($\mathrm{min}_r\, v_r = 0.93$)}, xmin=14.4, xmax=165.6, ymin=0.0646427, ymax=6.41475, xtick={20,60,100,140}, ytickten={-1,0}]
  \addplot[name path=lo2, draw=none, forget plot] table[x=x, y=SM2min] {\aspfigdir/data/fig4_R4.dat};
  \addplot[name path=hi2, draw=none, forget plot] table[x=x, y=SM2max] {\aspfigdir/data/fig4_R4.dat};
  \addplot[sm2band, forget plot] fill between[of=hi2 and lo2];
  \draw[refline] ({rel axis cs:0,0} |- {axis cs:0,1}) -- ({rel axis cs:1,0} |- {axis cs:0,1});
  \addplot[sm3, forget plot] table[x=x, y=SM3] {\aspfigdir/data/fig4_R4.dat};
  \addplot[sm2best, forget plot] table[x=x, y=SM2best] {\aspfigdir/data/fig4_R4.dat};
  \addplot[sm1, forget plot] table[x=x, y=SM1] {\aspfigdir/data/fig4_R4.dat};
\end{groupplot}
\node[anchor=south, inner sep=0pt] at ($(grp c1r1.outer north)!0.5!(grp c3r1.outer north)+(0,0pt)$)
  {\pgfplotslegendfromname{legfig4}};
\end{tikzpicture}
	\caption{Effect of the number of layers. Reconstruction error on
	misaligned multigraphs with the sensor-graph Fourier basis and
	$R=2,3,4$ layers, for $K_{\omega}\approx 0.08N$,
	$20\leq\vert\ccalS\vert-K_{\omega}\leq 160$, and
	$\mathrm{SNR}=15\,\mathrm{dB}$.}
	\label{fig_exp_layers}
\end{figure*}

\begin{figure*}[t]
	\centering
\providecommand{\aspfigdir}{./figures/experiments}
\ifcsname pgfplotsgroupplotsmath@decrease\endcsname\else\PackageError{asp figures}{pgfplots libraries not loaded}{Add to the PREAMBLE of the main file: \string\usepackage{pgfplots} \string\usepgfplotslibrary{groupplots,fillbetween} \string\usetikzlibrary{calc} \string\pgfplotsset{compat=1.18}}\fi
\ifdefined\aspstyleloaded\else\input{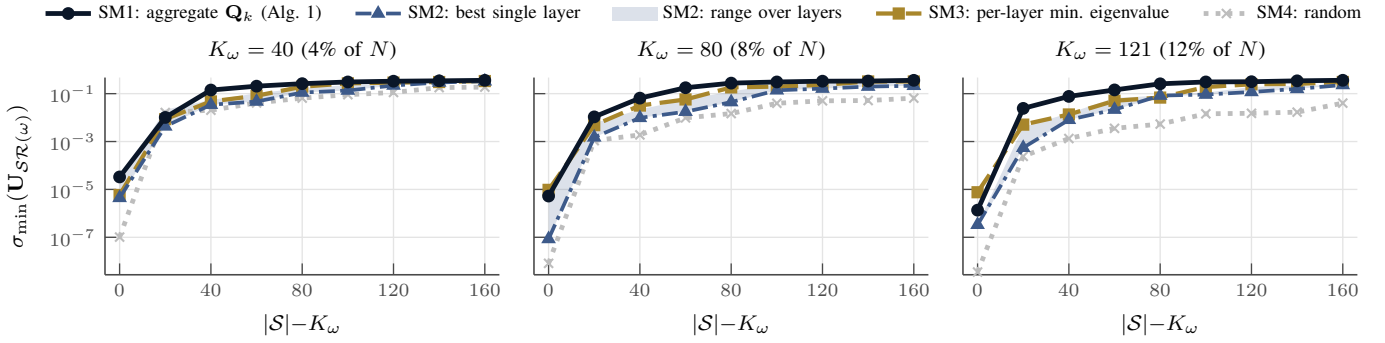}\fi
\begin{tikzpicture}
\pgfmathsetlengthmacro{\aspPW}{0.28*\textwidth}
\pgfmathsetlengthmacro{\aspPH}{\aspPW/2}
\begin{groupplot}[
  group style={group name=grp, group size=3 by 1, horizontal sep=0.18in, vertical sep=0.3in, xlabels at=edge bottom, xticklabels at=edge bottom, ylabels at=edge left, yticklabels at=edge left},
  asp axis, asp log, asp legend, width=\aspPW, height=\aspPH,
  xlabel={$\vert\mathcal{S}\vert - K_\omega$}, ylabel={$\sigma_{\min}(\mathbf{U}_{\mathcal{S}\mathcal{R}(\omega)})$},
]
  \nextgroupplot[title={$K_\omega = 40$ (4\% of $N$)}, xmin=-6.4, xmax=166.4, ymin=2.72367e-09, ymax=0.47023, xtick={0,40,80,120,160}, ytickten={-7,-5,-3,-1}, legend to name=legfig5]
  \addlegendimage{sm1}\addlegendentry{SM1: aggregate $\mathbf{Q}_k$ (Alg.~1)}
  \addlegendimage{sm2best}\addlegendentry{SM2: best single layer}
  \addlegendimage{draw=none, legend image code/.code={\path[draw=none, fill=asplayer, fill opacity=0.18] (0cm,-0.09cm) rectangle (0.6cm,0.09cm);}}\addlegendentry{SM2: range over layers}
  \addlegendimage{sm3}\addlegendentry{SM3: per-layer min.\ eigenvalue}
  \addlegendimage{sm4}\addlegendentry{SM4: random}
  \addplot[name path=lo0, draw=none, forget plot] table[x=x, y=SM2min] {\aspfigdir/data/fig5_bw4.dat};
  \addplot[name path=hi0, draw=none, forget plot] table[x=x, y=SM2max] {\aspfigdir/data/fig5_bw4.dat};
  \addplot[sm2band, forget plot] fill between[of=hi0 and lo0];
  \addplot[sm4, forget plot] table[x=x, y=SM4] {\aspfigdir/data/fig5_bw4.dat};
  \addplot[sm3, forget plot] table[x=x, y=SM3] {\aspfigdir/data/fig5_bw4.dat};
  \addplot[sm2best, forget plot] table[x=x, y=SM2best] {\aspfigdir/data/fig5_bw4.dat};
  \addplot[sm1, forget plot] table[x=x, y=SM1] {\aspfigdir/data/fig5_bw4.dat};
  \nextgroupplot[title={$K_\omega = 80$ (8\% of $N$)}, xmin=-6.4, xmax=166.4, ymin=2.72367e-09, ymax=0.47023, xtick={0,40,80,120,160}, ytickten={-7,-5,-3,-1}]
  \addplot[name path=lo1, draw=none, forget plot] table[x=x, y=SM2min] {\aspfigdir/data/fig5_bw8.dat};
  \addplot[name path=hi1, draw=none, forget plot] table[x=x, y=SM2max] {\aspfigdir/data/fig5_bw8.dat};
  \addplot[sm2band, forget plot] fill between[of=hi1 and lo1];
  \addplot[sm4, forget plot] table[x=x, y=SM4] {\aspfigdir/data/fig5_bw8.dat};
  \addplot[sm3, forget plot] table[x=x, y=SM3] {\aspfigdir/data/fig5_bw8.dat};
  \addplot[sm2best, forget plot] table[x=x, y=SM2best] {\aspfigdir/data/fig5_bw8.dat};
  \addplot[sm1, forget plot] table[x=x, y=SM1] {\aspfigdir/data/fig5_bw8.dat};
  \nextgroupplot[title={$K_\omega = 121$ (12\% of $N$)}, xmin=-6.4, xmax=166.4, ymin=2.72367e-09, ymax=0.47023, xtick={0,40,80,120,160}, ytickten={-7,-5,-3,-1}]
  \addplot[name path=lo2, draw=none, forget plot] table[x=x, y=SM2min] {\aspfigdir/data/fig5_bw12.dat};
  \addplot[name path=hi2, draw=none, forget plot] table[x=x, y=SM2max] {\aspfigdir/data/fig5_bw12.dat};
  \addplot[sm2band, forget plot] fill between[of=hi2 and lo2];
  \addplot[sm4, forget plot] table[x=x, y=SM4] {\aspfigdir/data/fig5_bw12.dat};
  \addplot[sm3, forget plot] table[x=x, y=SM3] {\aspfigdir/data/fig5_bw12.dat};
  \addplot[sm2best, forget plot] table[x=x, y=SM2best] {\aspfigdir/data/fig5_bw12.dat};
  \addplot[sm1, forget plot] table[x=x, y=SM1] {\aspfigdir/data/fig5_bw12.dat};
\end{groupplot}
\node[anchor=south, inner sep=0pt] at ($(grp c1r1.outer north)!0.5!(grp c3r1.outer north)+(0,0pt)$)
  {\pgfplotslegendfromname{legfig5}};
\end{tikzpicture}
	\caption{Conditioning of the sampling sets: smallest singular value of
	the sampled basis $\bbU_{\ccalS,\ccalR(\omega)}$ for the misaligned
	multigraphs with the sensor-graph Fourier basis and $R=2$ layers, for the
	three bandwidths (median over $5$ realizations). Larger values
	correspond to a smaller amplification of the noise in the
	reconstruction~\eqref{eq_exp_reconstruction}.}
	\label{fig_exp_sigma_min}
\end{figure*}

\begin{table*}[t]
	\vspace{10pt}
	\centering
	\captionsetup{belowskip=4pt}
	\caption{Reconstruction error $E$ at $\vert\ccalS\vert=K_{\omega}+20$
	for $K_{\omega}\approx 0.08N$ (average over $50$ signals, median over
	$5$ realizations), for every Fourier basis, number of layers, regime, and
	noise level. SM2
	(best) and SM2 (worst) are the best and the worst single layer at this
	sampling set size. The smallest error among SM1, SM2 (best), and SM3 is
	shown in bold.}
	\label{tab_exp_errors}
	\resizebox{\textwidth}{!}{
\begin{tabular}{lclcccccccccc}
\hline
 & & & \multicolumn{5}{c}{SNR $= 15$ dB} & \multicolumn{5}{c}{SNR $= 20$ dB} \\
Basis & $R$ & Regime & SM1 & SM2 (best) & SM2 (worst) & SM3 & SM4 & SM1 & SM2 (best) & SM2 (worst) & SM3 & SM4 \\
\hline
Sensor & 2 & aligned & 0.16 & 0.16 & 0.16 & \textbf{0.15} & 6.21 & 0.09 & 0.09 & 0.09 & \textbf{0.09} & 3.65 \\
 & 2 & misaligned & \textbf{0.50} & 1.03 & 4.19 & 1.14 & 6.21 & \textbf{0.29} & 0.59 & 2.45 & 0.66 & 3.65 \\
 & 3 & aligned & 0.17 & \textbf{0.16} & 0.16 & 0.16 & 4.60 & 0.09 & \textbf{0.09} & 0.09 & 0.09 & 2.47 \\
 & 3 & misaligned & \textbf{0.21} & 1.01 & 6.82 & 0.82 & 4.60 & \textbf{0.12} & 0.58 & 4.47 & 0.47 & 2.47 \\
 & 4 & aligned & 0.16 & 0.16 & 0.16 & \textbf{0.15} & 7.43 & 0.09 & 0.09 & 0.09 & \textbf{0.09} & 5.38 \\
 & 4 & misaligned & \textbf{0.19} & 2.18 & 2.68 & 4.01 & 7.43 & \textbf{0.11} & 1.28 & 1.73 & 1.98 & 5.38 \\
\hline
Swiss roll & 2 & aligned & \textbf{0.15} & 0.16 & 0.16 & 0.16 & $2.1\!\times\!10^{5}$ & \textbf{0.08} & 0.09 & 0.09 & 0.09 & $1.2\!\times\!10^{5}$ \\
 & 2 & misaligned & \textbf{1.37} & 15.56 & 19.76 & 3.51 & $2.1\!\times\!10^{5}$ & \textbf{0.78} & 8.90 & 10.76 & 1.91 & $1.2\!\times\!10^{5}$ \\
 & 3 & aligned & \textbf{0.26} & 0.43 & 0.75 & 0.28 & $2.4\!\times\!10^{4}$ & \textbf{0.14} & 0.25 & 0.38 & 0.17 & $1.2\!\times\!10^{4}$ \\
 & 3 & misaligned & \textbf{2.38} & 21.69 & 52.83 & 9.00 & $2.4\!\times\!10^{4}$ & \textbf{1.18} & 16.63 & 34.15 & 4.92 & $1.2\!\times\!10^{4}$ \\
 & 4 & aligned & 0.18 & \textbf{0.16} & 0.18 & 0.16 & $3.2\!\times\!10^{5}$ & 0.10 & \textbf{0.09} & 0.10 & 0.09 & $1.6\!\times\!10^{5}$ \\
 & 4 & misaligned & \textbf{3.88} & 19.68 & $1.3\!\times\!10^{3}$ & 18.33 & $3.2\!\times\!10^{5}$ & \textbf{1.53} & 10.22 & $6.3\!\times\!10^{2}$ & 9.82 & $1.6\!\times\!10^{5}$ \\
\hline
Dragon & 2 & aligned & 0.17 & \textbf{0.17} & 0.17 & 0.17 & 0.78 & \textbf{0.09} & 0.09 & 0.10 & 0.10 & 0.44 \\
 & 2 & misaligned & \textbf{0.23} & 0.35 & 0.52 & 0.31 & 0.78 & \textbf{0.13} & 0.19 & 0.30 & 0.18 & 0.44 \\
 & 3 & aligned & \textbf{0.17} & 0.17 & 0.18 & 0.17 & 1.31 & \textbf{0.10} & 0.10 & 0.10 & 0.10 & 0.71 \\
 & 3 & misaligned & \textbf{0.19} & 0.31 & 0.68 & 0.26 & 1.31 & \textbf{0.11} & 0.18 & 0.38 & 0.15 & 0.71 \\
 & 4 & aligned & 0.17 & \textbf{0.17} & 0.18 & 0.17 & 0.82 & 0.10 & \textbf{0.10} & 0.10 & 0.10 & 0.46 \\
 & 4 & misaligned & \textbf{0.19} & 0.38 & 0.64 & 0.25 & 0.82 & \textbf{0.10} & 0.21 & 0.40 & 0.14 & 0.46 \\
\hline
\end{tabular}
}
\end{table*}


\subsection{The Multigraphs}
\label{subsec_exp_multigraphs}

We consider multigraphs with $N=1000$ nodes and $R\in\{2,3,4\}$ layers,
whose shift operators are synthesized directly from their joint block
diagonalization, which gives full control over the invariant subspaces and the frequency
labels. The Fourier basis $\bbU$ is the Laplacian eigenbasis of one of three
graphs: a random geometric sensor graph and a Swiss-roll graph, generated
with PyGSP~\cite{pygsp}, and a connected nearest-neighbor graph built on $1000$
points of the Stanford dragon~\cite{curless1996volumetric,stanford3drep}. The block sizes $q_{\ell}$ are drawn
uniformly from $\{2,\ldots,10\}$, which results in $L\approx 160$ blocks
shared by all the layers, and each block $\mathbf{\Xi}_{\ell}^{(r)}$ is a
dense symmetric matrix with a random orthonormal eigenbasis and prescribed
eigenvalue magnitudes. We consider two regimes. In \emph{aligned}
multigraphs, the eigenvalue magnitudes of every layer increase with $\ell$
and are separated between consecutive blocks, so that every shift operator
is admissible in the sense of Definition~\ref{def_admissible}. In
\emph{misaligned} multigraphs, each block is dominated by a randomly chosen
layer, which fixes its label $\mu_{\ell}$, while the remaining layers
assign it smaller and widely spread magnitudes, so that the
ordering~\eqref{eq_freq_ordering} holds but admissibility fails on every
layer. We quantify admissibility by the fraction $v_{r}\in[0,1]$ of
consecutive block pairs whose spectral intervals overlap on layer $r$,
cf.~\eqref{eq_exp_violation}, so that $\bbS_{r}$ is admissible if and only
if $v_{r}=0$. Every aligned realization has $v_{r}=0$ on all layers, and
every misaligned realization has $\min_{r}v_{r}\geq 0.92$. The full
construction is described in Appendix~\ref{app_multigraph_generation}. We
emphasize that the uniqueness guarantee of
Theorem~\ref{thm_unique_aggregate}, on which
Algorithm~\ref{alg_multigraph_sampling} relies, does not require
admissibility, and therefore applies to both regimes.


\subsection{Signals and Bandwidth Models}
\label{subsec_exp_signals}

As discussed in Section~\ref{sec_multigraph_sp}, the dimension of
$\mathcal{PW}_{\omega}(\ccalM)$ can only take the values
$\{0,\,q_{1},\,q_{1}+q_{2},\ldots,N\}$. Hence, rather than prescribing the
bandwidth $\omega$ directly, we prescribe the number of Fourier components
$K_{\omega}=\vert\ccalR(\omega)\vert=\dim\mathcal{PW}_{\omega}(\ccalM)$. For
each target value in $\{0.04N,\,0.08N,\,0.12N\}=\{40,\,80,\,120\}$, we set
$K_{\omega}$ to the block boundary $\sum_{j\leq\ell}q_{j}$ closest to the
target. This guarantees that the bandlimited space consists of whole
invariant subspaces, i.e., that no block is truncated, in agreement with
Definition~\ref{def_bandlimited_signal}.

The signals are generated in the spectral domain. We draw a vector
$\widehat{\bbx}\in\mbR^{N}$ whose first $K_{\omega}$ entries are independent
standard Gaussian random variables and whose remaining entries are zero, and
we obtain the node-domain signal through the inverse multigraph Fourier
transform $\bbx=\bbU\widehat{\bbx}$. The resulting multigraph signal
$(\ccalM,\bbx)$ belongs to $\mathcal{PW}_{\omega}(\ccalM)$, and it is
normalized so that $\Vert\bbx\Vert_{2}=1$.

\subsection{Sampling Methods}
\label{subsec_exp_methods}

\noindent We use the following methods for selecting the sampling set $\ccalS$:
\begin{itemize}
\item \textbf{SM1 (aggregate):} Algorithm~\ref{alg_multigraph_sampling},
with order $k=1$.
\item \textbf{SM2 (single layer):} the method
of~\cite{ortega_proxies} with the same order $k=1$, applied to the shift
operator of a single layer $\bbS_{r}$, ignoring the remaining layers. We
run this method for every $r=1,\ldots,R$. In the figures, we report the
range of errors over the layers together with the error of the best single
layer. 
\item \textbf{SM3 (per-layer worst case):} a greedy rule driven by the
constant $\Lambda^{\min}$ of Definition~\ref{def_remov_set}. At each
iteration, the smallest eigenpair of
$\left[\bbS_{r}^{\top}\bbS_{r}\right]_{\ccalS^{c},\ccalS^{c}}$ is computed for every
layer $r$, the layer attaining the smallest eigenvalue is selected, and the
node where the corresponding eigenvector has the largest magnitude is added
to $\ccalS$. Notice that the layer driving the selection may change between iterations.
\item \textbf{SM4 (random):} $\ccalS$ is drawn uniformly at random, without
replacement, from $\ccalV$.
\end{itemize}
All the greedy methods start from $\ccalS=\emptyset$.

\subsection{Experiments}
\label{subsec_exp_protocol}

The samples are corrupted by additive white Gaussian noise, i.e., we observe
$\widetilde{\bbx}_{\ccalS}=\bbx_{\ccalS}+\bbn_{\ccalS}$, where the entries
of $\bbn_{\ccalS}\in\mbR^{\vert\ccalS\vert}$ are independent zero-mean
Gaussian random variables whose variance is chosen so that the
$\mathrm{SNR}$ equals $15\,\mathrm{dB}$ or $20\,\mathrm{dB}$. The two noise levels
correspond to two separate sets of experiments. The signal is reconstructed according to
\begin{equation}\label{eq_exp_reconstruction}
\bbx_{\mathrm{rec}}
   =
   \bbU_{\ccalR(\omega)}
   \left(
       \bbU_{\ccalS,\ccalR(\omega)}
   \right)^{\dagger}
   \widetilde{\bbx}_{\ccalS}
   ,
\end{equation}
where $(\cdot)^{\dagger}$ denotes the Moore--Penrose
pseudoinverse. 

For each multigraph, bandwidth, noise level, and sampling method, the
sampling set size $\vert\ccalS\vert$ ranges from $K_{\omega}$ to
$K_{\omega}+400$ in increments of $20$. Each experiment consists of $5$
independent realizations. In each realization, a new multigraph is
generated, i.e., a new partition, new blocks and, for the sensor and
Swiss-roll graphs, a new Fourier basis $\bbU$. The point cloud is
fixed, and so is its basis. 

Since the partition changes between realizations, so does the block
boundary closest to each target bandwidth, and $K_{\omega}$ varies by a
few units around its target value, e.g., between $76$ and $84$ for
$K_{\omega}\approx 0.08N$. The results of different realizations are
therefore aligned by the number of samples beyond the bandwidth,
$\vert\ccalS\vert-K_{\omega}$, and the value of $K_{\omega}$ reported in
the titles of the figures is the median over the realizations.

Within a realization, the sampling sets of
every method are computed once, $50$ bandlimited signals are drawn for
each bandwidth, and every signal receives one noise realization per noise
level, which is shared by all the sampling methods and set sizes. Hence,
all the methods are compared on exactly the same signals and the same
noise. Moreover, for a given number of layers $R$, the aligned and the
misaligned multigraphs of a realization share the same partition and the
same Fourier basis, and differ only in their blocks. For each realization, we
average the error over the $50$ signals, and we report the median of these
averages over the $5$ realizations. We use the median rather than the mean
because, in a few realizations, a method selects a set that is not a
uniqueness set, or is numerically indistinguishable from one, in which case
the least squares reconstruction amplifies the noise by a large factor, and a single such realization would dominate a mean
over five. In addition to the reconstruction error, we report the smallest singular
value $\sigma_{\min}(\bbU_{\ccalS,\ccalR(\omega)})$ of the sampled basis,
which measures the conditioning of the reconstruction
problem~\eqref{eq_exp_reconstruction}.

\subsection{Results}
\label{subsec_exp_results}


\noindent\textbf{Aligned versus misaligned multigraphs.}
Fig.~\ref{fig_exp_regimes} compares the two regimes on the same partition
and the same Fourier basis, and Fig.~\ref{fig_exp_regimes_zoom} shows the
relevant range in detail. On the aligned multigraph, where every layer is
admissible, SM1, the single layers, and SM3 are indistinguishable for
$\vert\ccalS\vert\geq K_{\omega}+20$, in agreement with
Corollary~\ref{cor_coupled_opt}. When the layers order the invariant
subspaces consistently, any of them certifies the optimal sampling sets. On the
misaligned multigraph, where no layer is admissible, the methods separate.
With only $20$ samples beyond the bandwidth, SM1 already provides a
meaningful reconstruction, whereas the best single layer and SM3 do not
reach errors below $1$ until more samples are added.

\noindent\textbf{Robustness across Fourier bases and bandwidths.}
Fig.~\ref{fig_exp_bases} shows that this behavior is consistent across the
three Fourier bases and the three bandwidths. Over all the misaligned
configurations, SM1 attains the smallest error among SM1, SM2, and SM3 in
$52$ of the $54$ combinations of basis, number of layers, bandwidth, and
noise level at $\vert\ccalS\vert=K_{\omega}+20$, and in all of them at
$\vert\ccalS\vert=K_{\omega}+40$ and $\vert\ccalS\vert=K_{\omega}+60$. In
median, the error of the best single layer, which is selected
\emph{a posteriori}, is $3.7$, $1.8$, and $1.3$ times larger than that of
SM1 at these three sizes, and the error of SM3 is $2.4$, $1.6$, and $1.2$
times larger. Table~\ref{tab_exp_errors} shows
that, for $K_{\omega}\approx0.08N$ and $\vert\ccalS\vert=K_{\omega}+20$, SM1
attains the smallest error in all the misaligned configurations and at both
noise levels.

\noindent\textbf{Effect of the number of layers.}
Fig.~\ref{fig_exp_layers} shows the misaligned multigraphs with
$R=2,3,4$ layers. As $R$ grows, each layer dominates a smaller fraction of
the blocks, so the information relevant for sampling becomes more
fragmented across the layers, and the advantage of SM1 grows accordingly.
Over all the misaligned configurations at $\vert\ccalS\vert=K_{\omega}+20$,
the median ratio between the error of the best single layer and that of
SM1 is $1.9$, $4.2$, and $4.2$ for $R=2,3,4$, and the corresponding ratio
for SM3 is $2.3$, $2.5$, and $2.9$. The median error of SM1
itself decreases with $R$, from $0.39$ for $R=2$ to $0.21$ for $R=3$ and
$0.20$ for $R=4$, whereas the median errors of the best single layer
($0.76$, $0.78$, and $1.23$) and of SM3 ($0.70$, $0.51$, and $0.68$) do
not decrease. That is, the additional layers are a source of
information for the aggregate operator, while they only dilute the
information available to each individual layer.

\noindent\textbf{Conditioning of the sampling sets.}
The source of these differences is explained by
Fig.~\ref{fig_exp_sigma_min}, which shows the smallest singular value of
the sampled basis. Over all the misaligned configurations, SM1 selects the
best-conditioned set in $96\%$ of the cases at
$\vert\ccalS\vert=K_{\omega}+20$ and in all of them at $K_{\omega}+40$ and
$K_{\omega}+60$. At $\vert\ccalS\vert=K_{\omega}+20$,
$\sigma_{\min}(\bbU_{\ccalS,\ccalR(\omega)})$ for SM1 is, in median, $3.0$
times larger than the largest value attained by a single layer and $2.8$
times larger than for SM3. We emphasize that, in the misaligned regime, the
aggregate operator $\mathbf{Q}_{1}$ is itself not admissible: its violation
fraction~\eqref{eq_exp_violation}, computed from the blocks
in~\eqref{eq_Qk_blocks}, is at least $0.96$ in every realization. The
advantage of SM1 is therefore not a consequence of
Corollary~\ref{cor_coupled_opt}, but of the unconditional guarantee of
Theorem~\ref{thm_unique_aggregate} and of the pooling of the separation of
all the layers quantified in Proposition~\ref{prop_domination}, whereas
the per-layer worst case of SM3 is limited at every step by the least
informative layer -- see Remark~\ref{rem_max_vs_min}.




\section{Conclusions}
\label{sec_conclusions}

We studied the sampling of bandlimited signals on multigraphs, where the
shift operators of the layers do not commute and admit only a joint block
diagonalization. We introduced a layer-free notion of frequency, which
allowed us to define bandlimited multigraph signals in a way that is
independent of the layer used to measure the frequencies, and we showed
that the resulting Paley-Wiener spaces are invariant under the entire
algebra of convolutional filters and that their dimensions are quantized by
the sizes of the blocks. We extended the notions of removable and
uniqueness sets to multigraphs, and we showed that the removability of the
complement of a sampling set on a single layer is enough to guarantee
uniqueness. That is, one informative layer suffices, and the degeneracies
of individual layers do not limit the sampling capacity of the multigraph.

The central conclusion of our analysis is that the exact optimality of a
sampling set is a property of the common Fourier basis of the multigraph,
and not of any of its constituent graphs. The spectral reach characterizes
uniqueness sets exactly, and, as a consequence, every admissible element of
the filter algebra certifies the same optimal sampling sets. The layers
re-enter the picture only through the computable estimates of optimality,
and since admissibility cannot be verified without spectral information,
the choice of the operator on which to run a single-graph sampling
algorithm is unknown a priori. We showed that this obstruction is dissolved
by the aggregate operator $\mathbf{Q}_{k}$, which shares the common Fourier
basis by construction, pools the information of all the layers, and
certifies uniqueness unconditionally. This led to a greedy sampling
algorithm for multigraphs that does not require the computation of the
joint block diagonalization, and whose cost per iteration is comparable to
that of the single-graph algorithm of~\cite{ortega_proxies} on each
layer.

The numerical experiments are consistent with these results. When every
layer is admissible, the proposed algorithm and the single-layer methods
are essentially indistinguishable, as predicted by
Corollary~\ref{cor_coupled_opt}. When no layer is admissible, the proposed
algorithm attains the smallest reconstruction error in almost every
configuration, even when it is compared with the best single layer selected
\emph{a posteriori}. Its advantage is larger for multigraphs with more
layers, and it is explained by the better conditioning of the selected
sampling sets. 

Several questions remain open. First, the aggregate operator considered
here uses only the powers of the shift operators, while mixed words probe
the interaction between the layers and may lead to tighter estimates when
the layers are strongly noncommuting. The systematic choice of these words
is a natural next step. Second, our results assume an exact joint block
diagonalization, whereas the shift operators of real multigraphs are
generically irreducible, with a single invariant subspace of dimension $N$. Extending the theory to approximate joint block diagonalizations, where
signals are only approximately bandlimited, is essential to bring these
results to real data. Finally, the
extension of these results to sequences of multigraphs of growing size, in
the spirit of the analysis of sampling sets on graphons~\cite{gphon_samp}, would allow the comparison and transfer of sampling sets across multigraphs
with different numbers of nodes.


 \bibliography{bibliography}
 \bibliographystyle{IEEEtran}



\clearpage
\newpage
\begin{center}
{\LARGE \textbf{Supplementary Material}}\\[1em]
{\large Proofs and Derivations}
\end{center}
\vspace{1em}


 \appendices



\section{Proofs}\label{app_proofs}

\subsection{Proof of Lemma~\ref{lemma_pw_invariance}}
\label{proof_lemma_pw_invariance}

\begin{proof}
Let $\mathbf{x}\in\mathcal{PW}_{\omega}(\ccalM)$ and let
$\mathbf{y}=p(\mathbf{S}_{1},\ldots,\mathbf{S}_{R})\mathbf{x}$ for an
arbitrary convolutional filter $p$. By the spectral representation of the
filter, the Fourier coefficients of
$\mathbf{y}$ satisfy, block by block,
\begin{equation}
\widehat{\mathbf{y}}_{\ell}
   =
   p\!\left(
        \mathbf{\Xi}_{\ell}^{(1)},
        \ldots,
        \mathbf{\Xi}_{\ell}^{(R)}
    \right)
    \widehat{\mathbf{x}}_{\ell}
    ,
    \qquad
    \ell=1,\ldots,L
    .
\end{equation}
If $\ell$ is such that $\mu_{\ell}>\omega$, then
$\widehat{\mathbf{x}}_{\ell}=\mathbf{0}$ by
Definition~\ref{def_bandlimited_signal}, and therefore
$\widehat{\mathbf{y}}_{\ell}=p(\mathbf{\Xi}_{\ell}^{(1)},\ldots,
\mathbf{\Xi}_{\ell}^{(R)})\,\mathbf{0}=\mathbf{0}$. Hence $\mathbf{y}$
satisfies~\eqref{eq_def_bandlimited} and
$\mathbf{y}\in\mathcal{PW}_{\omega}(\ccalM)$. The statement for words
follows by taking $p$ equal to a single word. 
\end{proof}

\subsection{Proof of Theorem~\ref{thm_bernstein_ineq}}
\label{proof_thm_bernstein_ineq}

\begin{proof}
Since $\mathbf{U}$ is orthonormal, and using the block
structure~\eqref{eq_jbd},
\begin{equation}
\left\Vert
    \mathbf{S}_{r}\mathbf{x}
\right\Vert_{2}^{2}
   =
   \left\Vert
       \operatorname{diag}
          \left(
              \mathbf{\Xi}_{1}^{(r)},\ldots,\mathbf{\Xi}_{L}^{(r)}
          \right)
       \widehat{\mathbf{x}}
   \right\Vert_{2}^{2}
   =
   \sum_{\ell=1}^{L}
       \left\Vert
           \mathbf{\Xi}_{\ell}^{(r)}
           \widehat{\mathbf{x}}_{\ell}
       \right\Vert_{2}^{2}
       .
\end{equation}
Since $(\ccalM,\mathbf{x})$ is $\omega$-bandlimited, the sum runs only over
the blocks with $\mu_{\ell}\leq\omega$, and for any such block,
$\Vert\mathbf{\Xi}_{\ell}^{(r)}\widehat{\mathbf{x}}_{\ell}\Vert_{2}
\leq\sigma_{\max}(\mathbf{\Xi}_{\ell}^{(r)})
\Vert\widehat{\mathbf{x}}_{\ell}\Vert_{2}
\leq\mu_{\ell}\Vert\widehat{\mathbf{x}}_{\ell}\Vert_{2}
\leq\omega\Vert\widehat{\mathbf{x}}_{\ell}\Vert_{2}$,
where the second inequality uses the definition of the frequency
label~\eqref{eq_freq_label}. Therefore,
\begin{equation}
\left\Vert
    \mathbf{S}_{r}\mathbf{x}
\right\Vert_{2}^{2}
   \leq
   \omega^{2}
   \sum_{\ell\,:\,\mu_{\ell}\leq\omega}
       \left\Vert
           \widehat{\mathbf{x}}_{\ell}
       \right\Vert_{2}^{2}
   =
   \omega^{2}
   \left\Vert
       \mathbf{x}
   \right\Vert_{2}^{2}
   ,
\end{equation}
using again the orthonormality of $\mathbf{U}$. 
\end{proof}

\subsection{Proof of Corollary~\ref{cor_word_bernstein}}
\label{proof_cor_word_bernstein}

\begin{proof}
We proceed by induction on the length $k=\vert w\vert$ of the word. The
case $k=1$ is Theorem~\ref{thm_bernstein_ineq}. Assume the claim holds for
all words of length $k-1$, and let $w=(r_{k},\ldots,r_{1})$. Write
$\mathbf{z}=\mathbf{S}_{r_{k-1}}\cdots\mathbf{S}_{r_{1}}\mathbf{x}$, so that
$\mathbf{S}_{w}\mathbf{x}=\mathbf{S}_{r_{k}}\mathbf{z}$. By
Lemma~\ref{lemma_pw_invariance}, $\mathbf{z}\in
\mathcal{PW}_{\omega}(\ccalM)$, so Theorem~\ref{thm_bernstein_ineq} applies
to $\mathbf{z}$, and combined with the induction hypothesis,
\begin{equation}
\left\Vert
    \mathbf{S}_{w}\mathbf{x}
\right\Vert_{2}
   =
   \left\Vert
       \mathbf{S}_{r_{k}}\mathbf{z}
   \right\Vert_{2}
   \leq
   \omega
   \left\Vert
       \mathbf{z}
   \right\Vert_{2}
   \leq
   \omega\cdot\omega^{k-1}
   \left\Vert
       \mathbf{x}
   \right\Vert_{2}
   =
   \omega^{k}
   \left\Vert
       \mathbf{x}
   \right\Vert_{2}
   .
\end{equation}
\end{proof}

\subsection{Proof of Lemma~\ref{lemma_lambda_eig}}
\label{proof_lemma_lambda_eig}

\begin{proof}
Any $\mathbf{x}\in L_{2}(\ccalS)$ can be written as
$\mathbf{x}=\mathbf{I}_{\ccalV\ccalS}\,\mathbf{x}_{\ccalS}$ with
$\mathbf{x}_{\ccalS}\in\mbR^{\vert\ccalS\vert}$, and
$\Vert\mathbf{x}\Vert_{2}=\Vert\mathbf{x}_{\ccalS}\Vert_{2}$. Then
\begin{equation}
\left\Vert
    \mathbf{S}_{r}\mathbf{x}
\right\Vert_{2}^{2}
   =
   \mathbf{x}_{\ccalS}^{\top}\,
   \mathbf{I}_{\ccalS\ccalV}\,
   \mathbf{S}_{r}^{\top}\mathbf{S}_{r}\,
   \mathbf{I}_{\ccalV\ccalS}\,
   \mathbf{x}_{\ccalS}
   =
   \mathbf{x}_{\ccalS}^{\top}
   \left[
       \mathbf{S}_{r}^{\top}\mathbf{S}_{r}
   \right]_{\ccalS,\ccalS}
   \mathbf{x}_{\ccalS}
   ,
\end{equation}
so that the infimum in~\eqref{eq_def_remov_set_1} is the minimum of the
Rayleigh quotient of the symmetric positive semidefinite matrix
$\left[\mathbf{S}_{r}^{\top}\mathbf{S}_{r}\right]_{\ccalS,\ccalS}
=(\mathbf{S}_{r}\mathbf{I}_{\ccalV\ccalS})^{\top}
(\mathbf{S}_{r}\mathbf{I}_{\ccalV\ccalS})$, which is attained and equals its
smallest eigenvalue; this proves~\eqref{eq_lambda_eig}. Finally,
$\lambda_{\min}([\mathbf{S}_{r}^{\top}\mathbf{S}_{r}]_{\ccalS,\ccalS})>0$ if and
only if $\mathbf{S}_{r}\mathbf{I}_{\ccalV\ccalS}$ has full column rank, i.e.,
if and only if no nonzero $\mathbf{x}\in L_{2}(\ccalS)$ satisfies
$\mathbf{S}_{r}\mathbf{x}=\mathbf{0}$, which
is~\eqref{eq_kernel_condition}. 
\end{proof}

\subsection{Proof of Theorem~\ref{thm_unique_set}}
\label{proof_thm_unique_set}

\begin{proof}
Let $\mathbf{z}=\mathbf{x}-\mathbf{y}$. Since
$\mathcal{PW}_{\omega}(\ccalM)$ is a subspace (cf.~\eqref{eq_pw_as_range}),
$\mathbf{z}$ is $\omega$-bandlimited, and since
$\mathbf{x}\vert_{\ccalS}=\mathbf{y}\vert_{\ccalS}$, we have
$\mathbf{z}\in L_{2}(\ccalS^{c})$. Suppose, for the sake of contradiction,
that $\mathbf{z}\neq\mathbf{0}$, and let $r^{*}$ be an index attaining the
maximum in the definition of $\Lambda_{\ccalS^{c}}^{\max}$. By
Definition~\ref{def_remov_set} applied on the $\mathbf{S}_{r^{*}}$-axis,
\begin{equation}
\left\Vert
    \mathbf{S}_{r^{*}}\mathbf{z}
\right\Vert_{2}
   \geq
   \Lambda_{r^{*},\ccalS^{c}}
   \left\Vert
       \mathbf{z}
   \right\Vert_{2}
   =
   \Lambda_{\ccalS^{c}}^{\max}
   \left\Vert
       \mathbf{z}
   \right\Vert_{2}
   ,
\end{equation}
while the Bernstein inequality of Theorem~\ref{thm_bernstein_ineq} gives
$\Vert\mathbf{S}_{r^{*}}\mathbf{z}\Vert_{2}\leq
\omega\Vert\mathbf{z}\Vert_{2}$. Combining the two,
$\Lambda_{\ccalS^{c}}^{\max}\leq\omega$, which
contradicts~\eqref{eq_thm_unique_condition}. Hence
$\mathbf{z}=\mathbf{0}$. 
\end{proof}

\subsection{Proof of Theorem~\ref{thm_reach_characterization}}
\label{proof_thm_reach_characterization}

\begin{proof}
We first note the standard equivalence: $\ccalS$ is a uniqueness set for
$\mathcal{PW}_{\omega}(\ccalM)$ if and only if
\begin{equation}\label{eq_proof_reach_equiv}
\mathcal{PW}_{\omega}(\ccalM)
\cap
L_{2}(\ccalS^{c})
=
\{\mathbf{0}\}
.
\end{equation}
Indeed, if $\mathbf{z}\neq\mathbf{0}$ belongs to the intersection, then
$\mathbf{0}$ and $\mathbf{z}$ are distinct $\omega$-bandlimited signals
agreeing on $\ccalS$; conversely, if $\mathbf{x}\neq\mathbf{y}$ are
$\omega$-bandlimited and agree on $\ccalS$, then
$\mathbf{x}-\mathbf{y}$ is a nonzero element of the intersection.

Without loss of generality, we may assume that the labels
$\mu_{1}<\ldots<\mu_{L}$ are pairwise distinct: consecutive blocks with
equal labels can be merged into a single block, since the merged block
subspaces are invariant, and both $\mathcal{PW}_{\omega}(\ccalM)$
in~\eqref{eq_pw_as_range} and the thresholds in~\eqref{eq_thm_reach} depend
on the partition only through the distinct label values.

By~\eqref{eq_pw_as_range},
$\mathcal{PW}_{\omega}(\ccalM)=\bigoplus_{j\leq\ell(\omega)}
\operatorname{range}(\mathbf{U}_{j})$ with
$\ell(\omega)=\max\{\ell:\mu_{\ell}\leq\omega\}$. By the minimality in
Definition~\ref{def_spectral_reach}, the
intersection~\eqref{eq_proof_reach_equiv} is trivial for every
$\ell(\omega)<\ell^{*}(\ccalS)$ and nontrivial for every
$\ell(\omega)\geq\ell^{*}(\ccalS)$, since the filtration
$\bigoplus_{j\leq\ell}\operatorname{range}(\mathbf{U}_{j})$ is increasing in
$\ell$. Finally, because the labels are strictly increasing,
$\ell(\omega)\geq\ell^{*}(\ccalS)$ holds if and only if
$\mu_{\ell^{*}(\ccalS)}\leq\omega$. Therefore, $\ccalS$ is a uniqueness set
for $\mathcal{PW}_{\omega}(\ccalM)$ if and only if
$\omega<\mu_{\ell^{*}(\ccalS)}$, which proves~\eqref{eq_thm_reach}.

For~\eqref{eq_opt_reach}, note that the family of bandwidths for which
$\ccalS$ is a uniqueness set is the interval
$[0,\mu_{\ell^{*}(\ccalS)})$, which is increasing (with respect to
inclusion) in $\ell^{*}(\ccalS)$ since the labels are nondecreasing; when
the labels are pairwise distinct the map
$\ell\mapsto\mu_{\ell}$ is strictly increasing and the two $\argmax$ sets
in~\eqref{eq_opt_reach} coincide. 
\end{proof}

\subsection{Proof of Corollary~\ref{cor_coupled_opt}}
\label{proof_cor_coupled_opt}

\begin{proof}
Since $\mathbf{T}$ is symmetric and
$\mathbf{T}=\mathbf{U}\operatorname{diag}(\mathbf{\Theta}_{1},\ldots,
\mathbf{\Theta}_{L})\mathbf{U}^{\top}$ with $\mathbf{U}$ orthonormal, each
block $\mathbf{\Theta}_{\ell}$ is symmetric, so the eigendecompositions
$\mathbf{\Theta}_{\ell}=\mathbf{V}_{\ell}\mathbf{D}_{\ell}
\mathbf{V}_{\ell}^{\top}$ in the statement exist and~\eqref{eq_UT} holds. The
columns of $\mathbf{U}_{\mathbf{T}}$ associated with the $\ell$th block are
the columns of $\mathbf{U}_{\ell}\mathbf{V}_{\ell}$, which span
$\operatorname{range}(\mathbf{U}_{\ell})$, cf.~\eqref{eq_UT_blocks}. By
definition, $\omega_{\mathbf{T}}(\boldsymbol{\phi})$ equals the largest
$\vert\lambda\vert$ over the eigencomponents of $\boldsymbol{\phi}$ in the
basis $\mathbf{U}_{\mathbf{T}}$, and, by the separation
condition~\eqref{eq_def_admissible}, eigencomponents lying in block $\ell$
carry labels inside the $\ell$th spectral interval
$[\min_{\lambda\in\operatorname{spec}(\mathbf{\Theta}_{\ell})}
\vert\lambda\vert,\
\max_{\lambda\in\operatorname{spec}(\mathbf{\Theta}_{\ell})}
\vert\lambda\vert]$, and these intervals are pairwise disjoint and
increasing in $\ell$.

\emph{Proof of (i).} Write $\ell^{*}=\ell^{*}(\ccalS)$. For the upper
bound, by Definition~\ref{def_spectral_reach} there exists a nonzero
$\boldsymbol{\phi}_{0}\in L_{2}(\ccalS^{c})\cap
\bigoplus_{j\leq\ell^{*}}\operatorname{range}(\mathbf{U}_{j})$. All the
eigencomponents of $\boldsymbol{\phi}_{0}$ lie in blocks $j\leq\ell^{*}$,
so, by the ordering of the intervals,
$\omega_{\mathbf{T}}(\boldsymbol{\phi}_{0})\leq
\max_{\lambda\in\operatorname{spec}(\mathbf{\Theta}_{\ell^{*}})}
\vert\lambda\vert$, whence
$\nu_{\mathbf{T}}(\ccalS)\leq
\max_{\lambda\in\operatorname{spec}(\mathbf{\Theta}_{\ell^{*}})}
\vert\lambda\vert$. For the lower bound, let $\boldsymbol{\phi}\in
L_{2}(\ccalS^{c})$ be nonzero. If all the eigencomponents of
$\boldsymbol{\phi}$ were contained in blocks $j\leq\ell^{*}-1$, then
$\boldsymbol{\phi}$ would be a nonzero element of
$L_{2}(\ccalS^{c})\cap\bigoplus_{j\leq\ell^{*}-1}
\operatorname{range}(\mathbf{U}_{j})$, contradicting the minimality of
$\ell^{*}$. Hence $\boldsymbol{\phi}$ has an eigencomponent in some block
$j\geq\ell^{*}$, and, by the ordering of the intervals,
$\omega_{\mathbf{T}}(\boldsymbol{\phi})\geq
\min_{\lambda\in\operatorname{spec}(\mathbf{\Theta}_{\ell^{*}})}
\vert\lambda\vert$. Taking the minimum over $\boldsymbol{\phi}$
proves~(i).

\emph{Proof of (ii).} If $\ell^{*}(\ccalS_{1})<\ell^{*}(\ccalS_{2})$, then,
using (i) twice and the disjointness and ordering of the intervals,
\begin{multline}
\nu_{\mathbf{T}}(\ccalS_{1})
\leq
\max_{\lambda\in\operatorname{spec}
(\mathbf{\Theta}_{\ell^{*}(\ccalS_{1})})}\vert\lambda\vert
<
\min_{\lambda\in\operatorname{spec}
(\mathbf{\Theta}_{\ell^{*}(\ccalS_{1})+1})}\vert\lambda\vert
\\
\leq
\min_{\lambda\in\operatorname{spec}
(\mathbf{\Theta}_{\ell^{*}(\ccalS_{2})})}\vert\lambda\vert
\leq
\nu_{\mathbf{T}}(\ccalS_{2})
.
\end{multline}

\emph{Proof of (iii).} Let $\ccalS^{*}$ maximize
$\nu_{\mathbf{T}}(\cdot)$ over the sets of size $m$, and suppose some
$\ccalS'$ of size $m$ satisfies
$\ell^{*}(\ccalS')>\ell^{*}(\ccalS^{*})$. Then, by (ii),
$\nu_{\mathbf{T}}(\ccalS')>\nu_{\mathbf{T}}(\ccalS^{*})$, a contradiction.
Hence $\ccalS^{*}$ maximizes $\ell^{*}(\cdot)$, and, by
Theorem~\ref{thm_reach_characterization}, it is an optimal sampling set of
the multigraph.
\end{proof}

\subsection{Proof of Theorem~\ref{thm_unique_aggregate}}
\label{proof_thm_unique_aggregate}

\begin{proof}
Let $\mathbf{z}=\mathbf{x}-\mathbf{y}$. As in the proof of
Theorem~\ref{thm_unique_set}, $\mathbf{z}\in
\mathcal{PW}_{\omega}(\ccalM)\cap L_{2}(\ccalS^{c})$. Suppose
$\mathbf{z}\neq\mathbf{0}$. On the one hand, writing
$\mathbf{z}=\mathbf{I}_{\ccalV\ccalS^{c}}\mathbf{z}_{\ccalS^{c}}$ and using
the Rayleigh quotient of the symmetric matrix
$[\mathbf{Q}_{k}]_{\ccalS^{c},\ccalS^{c}}$,
\begin{multline}
\sum_{r=1}^{R}
    \left\Vert
        \mathbf{S}_{r}^{k}\mathbf{z}
    \right\Vert_{2}^{2}
=
\mathbf{z}_{\ccalS^{c}}^{\top}
\left[\mathbf{Q}_{k}\right]_{\ccalS^{c},\ccalS^{c}}
\mathbf{z}_{\ccalS^{c}}
\\
\geq
\lambda_{\min}
   \!\left(
       \left[\mathbf{Q}_{k}\right]_{\ccalS^{c},\ccalS^{c}}
    \right)
\left\Vert
    \mathbf{z}
\right\Vert_{2}^{2}
=
R\,
\Gamma_{k,\ccalS^{c}}^{2k}
\left\Vert
    \mathbf{z}
\right\Vert_{2}^{2}
.
\end{multline}
On the other hand, by Corollary~\ref{cor_word_bernstein} applied to the
words $w=(r,\ldots,r)$ of length $k$,
\begin{equation}
\sum_{r=1}^{R}
    \left\Vert
        \mathbf{S}_{r}^{k}\mathbf{z}
    \right\Vert_{2}^{2}
\leq
R\,
\omega^{2k}
\left\Vert
    \mathbf{z}
\right\Vert_{2}^{2}
.
\end{equation}
Combining the two displays and dividing by
$R\Vert\mathbf{z}\Vert_{2}^{2}>0$ yields
$\Gamma_{k,\ccalS^{c}}\leq\omega$, which
contradicts~\eqref{eq_thm_unique_aggregate}. Hence
$\mathbf{z}=\mathbf{0}$. The last claim follows since the argument holds
for each $k\in\ccalK$ separately, so the strongest of the sufficient
conditions applies.
\end{proof}

\subsection{Proof of Proposition~\ref{prop_domination}}
\label{proof_prop_domination}

\begin{proof}
Since the operation of taking principal submatrices is linear,
$[\mathbf{Q}_{1}]_{\ccalS,\ccalS}=\sum_{r=1}^{R}
[\mathbf{S}_{r}^{\top}\mathbf{S}_{r}]_{\ccalS,\ccalS}$, a sum of symmetric
positive semidefinite matrices. By Weyl's inequality,
\begin{equation}\label{eq_proof_dom_weyl}
\lambda_{\min}
   \!\left(
       \left[\mathbf{Q}_{1}\right]_{\ccalS,\ccalS}
    \right)
\;\geq\;
\sum_{r=1}^{R}
    \lambda_{\min}
       \!\left(
           \left[
               \mathbf{S}_{r}^{\top}\mathbf{S}_{r}
           \right]_{\ccalS,\ccalS}
        \right)
=
\sum_{r=1}^{R}
    \Lambda_{r,\ccalS}^{2}
    ,
\end{equation}
where the equality uses Lemma~\ref{lemma_lambda_eig}. On the one hand,
$\sum_{r}\Lambda_{r,\ccalS}^{2}\geq(\Lambda_{\ccalS}^{\max})^{2}$, which
gives $\Gamma_{1,\ccalS}=[\lambda_{\min}([\mathbf{Q}_{1}]_{\ccalS,\ccalS})/R]^{1/2}
\geq\Lambda_{\ccalS}^{\max}/\sqrt{R}$. On the other hand,
$\sum_{r}\Lambda_{r,\ccalS}^{2}\geq R\,(\Lambda_{\ccalS}^{\min})^{2}$,
which gives $\Gamma_{1,\ccalS}\geq\Lambda_{\ccalS}^{\min}$. If the layer
constants are not all equal, the latter inequality between the sums is
strict, and hence so is
$\Gamma_{1,\ccalS}>\Lambda_{\ccalS}^{\min}$.
\end{proof}

\subsection{Proof of Proposition~\ref{prop_monotonicity}}
\label{proof_prop_monotonicity}

\begin{proof}
If $\ccalS_{1}\subseteq\ccalS_{2}$, then
$\ccalS_{2}^{c}\subseteq\ccalS_{1}^{c}$, and
$[\mathbf{Q}_{k}]_{\ccalS_{2}^{c}}$ is a principal submatrix of the
symmetric matrix $[\mathbf{Q}_{k}]_{\ccalS_{1}^{c}}$. By the Cauchy
eigenvalue interlacing theorem~\cite[Thm.~4.3.17]{horn2012matrix}, the smallest eigenvalue can only
increase (or remain unchanged) when passing to a principal submatrix,
\begin{equation}
\lambda_{\min}
   \!\left(
       \left[\mathbf{Q}_{k}\right]_{\ccalS_{1}^{c}}
    \right)
\;\leq\;
\lambda_{\min}
   \!\left(
       \left[\mathbf{Q}_{k}\right]_{\ccalS_{2}^{c}}
    \right)
    ,
\end{equation}
and the claim follows by taking $2k$-th roots and dividing by $R$ inside
the root, both of which preserve the inequality.
\end{proof}


\subsection{Additional Details on The Multigraphs used in the Experiments}
\label{app_multigraph_generation}

We consider multigraphs $\ccalM = \left( \ccalV,
\{\ccalE_r,w_{r}\}_{r=1}^{R} \right)$ with $N=\vert\ccalV\vert=1000$ nodes
and $R\in\{2,3,4\}$ layers, with shift operators $\{ \bbS_{r}
\}_{r=1}^{R}$. To have full control over the block diagonal structure, the
shift operators are synthesized directly from their joint block
diagonalization~\eqref{eq_jbd}, where the orthonormal matrix $\bbU\in\mbR^{N\times N}$ and the blocks
$\mathbf{\Xi}_\ell^{(r)}$ are generated as follows.

\noindent\textbf{Fourier basis.} The matrix $\bbU$ is the eigenvector matrix
of the combinatorial Laplacian of an undirected graph on $N$ nodes, with its
columns sorted by increasing Laplacian eigenvalue. We consider three such graphs:
\begin{enumerate}
    \item a random geometric sensor graph, generated with the PyGSP
    toolbox~\cite{pygsp}.
    \item a Swiss-roll manifold graph, generated with the PyGSP
    toolbox~\cite{pygsp}.
    \item a graph built on the Stanford
    dragon~\cite{curless1996volumetric,stanford3drep}, from which $N=1000$
    points are selected by farthest-point sampling. Each point is connected
    to its $11$ nearest neighbors and the graph is symmetrized; the weight of
    the edge between $\mathbf{p}_{i}$ and $\mathbf{p}_{j}$ is
    $\exp(-\Vert\mathbf{p}_{i}-\mathbf{p}_{j}\Vert_{2}^{2}/(s_{i}s_{j}))$,
    where $s_{i}$ is the distance from $\mathbf{p}_{i}$ to its $11$th
    nearest neighbor~\cite{zelnik2004self}. The resulting graph is connected,
    with average degree $12.3$.
\end{enumerate}

\noindent\textbf{Partition.} The block sizes $q_{\ell}$ of the partition
$\tau_{N}=(q_{1},\ldots,q_{L})$ are drawn uniformly at random from
$\{2,\ldots,10\}$, subject to $\sum_{\ell=1}^{L}q_{\ell}=N$, which results
in $L\approx 160$ blocks. The same partition is shared by all the layers,
as required by~\eqref{eq_jbd}.

\noindent\textbf{Blocks.} Each block $\mathbf{\Xi}_\ell^{(r)}$ is a dense
symmetric matrix of the form
\begin{equation}\label{eq_exp_block}
\mathbf{\Xi}_\ell^{(r)}
   =
   \mathbf{O}_{\ell}^{(r)}
   \operatorname{diag}\!\left(
       \lambda_{\ell,1}^{(r)},\ldots,\lambda_{\ell,q_{\ell}}^{(r)}
   \right)
   \left(\mathbf{O}_{\ell}^{(r)}\right)^{\!\top},
\end{equation}
where $\mathbf{O}_{\ell}^{(r)}\in\mbR^{q_{\ell}\times q_{\ell}}$ is a
random orthonormal matrix drawn from the Haar distribution and the
eigenvalues $\lambda_{\ell,i}^{(r)}$ have random signs. Since
$\mathbf{\Xi}_\ell^{(r)}$ is symmetric, its singular values are the
magnitudes $\vert\lambda_{\ell,i}^{(r)}\vert$, so the eigenvalue magnitudes
directly control $\sigma_{\max}(\mathbf{\Xi}_\ell^{(r)})$ and, through it,
the labels $\mu_{\ell}$ in~\eqref{eq_freq_label}. Let
$t_{1}<t_{2}<\cdots<t_{L}$ be target magnitudes evenly spaced over
$[0.1,\,1.5]$, with gap $\delta=1.4/(L-1)$. In every layer, the eigenvalue
magnitudes of block $\ell$ are drawn independently and uniformly from
$[t_{\ell},\,t_{\ell}+\delta/4]$. Consequently,
\begin{equation}\label{eq_exp_aligned_separation}
\max_{\lambda\in\operatorname{spec}(\mathbf{\Xi}_{\ell}^{(r)})}
    \vert\lambda\vert
\;\leq\;
t_{\ell}+\frac{\delta}{4}
\;<\;
t_{\ell+1}
\;\leq\;
\min_{\lambda\in\operatorname{spec}(\mathbf{\Xi}_{\ell+1}^{(r)})}
    \vert\lambda\vert
\end{equation}
for every $\ell$ and $r$, i.e., every shift operator is admissible in the
sense of Definition~\ref{def_admissible} and, in particular,
$\sigma_{\max}(\mathbf{\Xi}_\ell^{(r)})$ is increasing in $\ell$ on every
layer, so that the ordering~\eqref{eq_freq_ordering} holds. Finally, each
$\bbS_r$ is normalized by its spectral radius, which rescales each layer
uniformly and preserves~\eqref{eq_exp_aligned_separation}. The layers
differ from one another through the random rotations
$\mathbf{O}_{\ell}^{(r)}$, the random signs, and the random magnitudes
within each cluster. We refer to the multigraphs obtained in this way as
\emph{aligned}. We note that the width of the clusters must be smaller than
the gap $\delta$ between consecutive targets, which is approximately
$9\times10^{-3}$ for the partitions considered here: with clusters of width
$0.05$, for instance, the spectral intervals of most consecutive blocks
overlap and the layers are no longer admissible.

\noindent\textbf{Aligned and misaligned multigraphs.} In the aligned regime,
Corollary~\ref{cor_coupled_opt} predicts that a single layer already
certifies the optimal sampling sets, and the sampling methods can only
differ through finite-order effects. To test the methods in the regime
where no individual layer can be trusted, we also generate
\emph{misaligned} multigraphs, in which the ordering~\eqref{eq_freq_ordering}
of the labels $\mu_{\ell}$ holds, but the admissibility
condition~\eqref{eq_def_admissible} fails on every layer. These are
generated as follows. Let $0<m_{1}<m_{2}<\cdots<m_{L}=1$ be target
magnitudes evenly spaced over $[0.1,\,1]$.
\begin{enumerate}
\item \emph{Dominant layers.} For each block $\ell$, a dominant layer
$r_{\ell}\in\{1,\ldots,R\}$ is drawn uniformly at random. The block
$\mathbf{\Xi}_{\ell}^{(r_{\ell})}$ is generated as
in~\eqref{eq_exp_block}, with the largest eigenvalue magnitude set exactly
to $m_{\ell}$.
\item \emph{Subdominant layers.} For every $r\neq r_{\ell}$, the block
$\mathbf{\Xi}_{\ell}^{(r)}$ is generated as in~\eqref{eq_exp_block} with
the largest eigenvalue magnitude equal to $\beta_{\ell}^{(r)}m_{\ell}$,
where $\beta_{\ell}^{(r)}$ is drawn uniformly from $[\beta_{0},\,1)$ for a
fixed $\beta_{0}\in(0,1)$.
\item \emph{Intra-block spread.} In all the layers, the remaining
eigenvalue magnitudes of a block are drawn uniformly from
$[\rho\,\sigma_{\max}(\mathbf{\Xi}_{\ell}^{(r)}),\,
\sigma_{\max}(\mathbf{\Xi}_{\ell}^{(r)})]$ for a fixed spread parameter
$\rho\in(0,1]$, again with random signs.
\end{enumerate}
By construction, $\mu_{\ell}=\max_{r}\sigma_{\max}(\mathbf{\Xi}_{\ell}^{(r)})
=m_{\ell}$, so the ordering~\eqref{eq_freq_ordering} holds strictly without
any regeneration step. Furthermore, $\max_{\ell}\mu_{\ell}=m_{L}=1$, so no
normalization of the shift operators is needed. We avoid normalizing each
$\bbS_{r}$ by its own spectral radius, since rescaling the layers by
different factors would change the labels $\mu_{\ell}$ and could destroy
the ordering. Admissibility of the individual layers, on the other hand,
fails for two independent reasons:
\begin{itemize}
\item On a layer $r$, $\sigma_{\max}(\mathbf{\Xi}_{\ell}^{(r)})$ is no
longer monotone in $\ell$: whenever $r$ dominates block $\ell-1$ but not
block $\ell$, we may have $\beta_{\ell}^{(r)}m_{\ell}<m_{\ell-1}$, which
happens for consecutive blocks as soon as $\beta_{0}<m_{\ell-1}/m_{\ell}$.
The parameter $\beta_{0}$ controls how severely each layer misorders the
blocks.
\item For $\rho<1$, the spectral intervals of consecutive blocks overlap
even on a layer where $\sigma_{\max}(\mathbf{\Xi}_{\ell}^{(r)})$ happens to
be monotone. This is because condition~\eqref{eq_def_admissible} compares
the \emph{smallest} magnitude of block $\ell+1$ with the largest magnitude
of block $\ell$, and the two intervals can overlap whenever
$\rho\,\sigma_{\max}(\mathbf{\Xi}_{\ell+1}^{(r)})<
\sigma_{\max}(\mathbf{\Xi}_{\ell}^{(r)})$.
\end{itemize}
Since the blocks are known in closed form, the admissibility of each layer
can be verified exactly. We quantify it by the \emph{violation fraction}
\begin{equation}\label{eq_exp_violation}
v_{r}
   =
   \frac{
   \left\vert
   \left\lbrace
       \ell
       \, : \,
       \sigma_{\max}\!\left(\mathbf{\Xi}_{\ell}^{(r)}\right)
       \geq
       \sigma_{\min}\!\left(\mathbf{\Xi}_{\ell+1}^{(r)}\right)
   \right\rbrace
   \right\vert
   }{L-1}
   \in[0,1]
   ,
\end{equation}
where, since the blocks are symmetric, $\sigma_{\max}$ and $\sigma_{\min}$
are the largest and smallest eigenvalue magnitudes, so that $\bbS_{r}$ is
admissible if and only if $v_{r}=0$. For the
misaligned multigraphs we use $\beta_{0}=0.3$ and $\rho=0.2$, and we report
$\min_{r}v_{r}$ for each realization, which certifies that none of the
layers is admissible. In all the misaligned realizations we obtain
$\min_{r}v_{r}\geq 0.92$, i.e., the separation condition fails for
at least $92\%$ of the consecutive block pairs on every layer, while
$v_{r}=0$ for every layer of every aligned realization.

We emphasize that the
uniqueness guarantee of Theorem~\ref{thm_unique_aggregate}, on which
Algorithm~\ref{alg_multigraph_sampling} relies, does not require
admissibility of any operator, and therefore applies equally to aligned and
misaligned multigraphs.

\end{document}